\documentclass{aa}
\usepackage{lastpage}
\usepackage{amssymb}
\usepackage{amsmath}
\usepackage{hyperref}
\usepackage{orcidlink}
\usepackage{float}
\usepackage[version=3]{mhchem}
\usepackage{subfigure}
\usepackage{graphicx}
\usepackage{makecell}
\usepackage{siunitx}
\usepackage{xcolor}
\usepackage[normalem]{ulem}
\DeclareSIUnit\erg{erg}
\DeclareSIUnit\Jys{\erg \centi \per \metre \squared \per \second}
 \DeclareSIUnit\angstrom{\text {Å}}

\title{Performance of photometric template fitting for ultra-high-redshift galaxies}

\begin{document}

\author{Thorbj\o rn Clausen\orcidlink{0009-0006-7165-3828}\inst{1,2} 
\and Charles L. Steinhardt\orcidlink{0000-0003-3780-6801}\inst{3,4} 
\and Arden Shao\orcidlink{0009-0002-6875-5111}\inst{5}
\and Gaurav Senthil Kumar\orcidlink{0000-0003-4454-351X}\inst{1,2}
}

\institute{Cosmic Dawn Center (DAWN), Denmark
\and Niels Bohr Institute, University of Copenhagen, Jagtvej 128, DK-2200, Copenhagen N, Denmark
\and Department of Physics and Astronomy, University of Missouri, Columbia, MO, USA
\and DARK Cosmology Center, University of Copenhagen, Jagtvej 128, DK-2200, Copenhagen N, Denmark
\and California Institute of Technology, 1200 E. California Blvd., Pasadena, CA 91125, USA
}

\date{Received 1 December 2024; accepted 17 March 2025}

\abstract{The James Webb Space Telescope (JWST) has enabled the discovery of a significant population of galaxies at $z > 10$. Our understanding of the astrophysical properties of these galaxies relies on fitting templates developed using models predicting the differences between these first galaxies and lower-redshift counterparts. In this work, tests are performed on several of these high-redshift template sets in order to determine how successful they are at predicting both photometric redshifts and full spectral energy distributions (SEDs). Our work shows that the best templates for photometric redshift estimation differ from the best templates for predicting the full SED. Overall, some templates perform adequately at photometric redshift estimation, while all are generally poor predictors of the full SED. A few objects in particular are poorly fit by all the template sets tested. We conclude that although photometric redshifts can be reliable when given a high enough observational depth and adequate filters, models are not yet able to produce robust astrophysical properties for these ultra-high-redshift galaxies.}

\maketitle

\section{Introduction}
\label{sec:intro}
Due to the high observational cost, large blind spectroscopic surveys are infeasible at very high redshift. As a result, all recent high-redshift galaxy studies have used either photometric surveys alone or spectroscopic follow-up of photometric targets. Consequently, our knowledge of the first galaxies relies on an ability to successfully identify high-redshift galaxies as well as several of their properties from photometry alone. This is done by fitting a series of spectral energy distribution (SED) models to the observed multi-wavelength photometry and then assuming that the properties of the best-fit model are also those of the galaxy.

However, there is a high computational complexity in fitting properties to photometry due to a combination of many astrophysical parameters and a search space with a large number of local minima that can lie far from the global minimum \citep{Speagle_2016}. In practice, solving this problem requires finding a way to significantly reduce the search space. One of the most successful techniques, which is employed by codes such as EAZY \citep{Brammer_2008}, is to choose a limited basis of templates and then to find the best-fit linear combination of the templates to the observed photometry.

The main drawback of this method is that success hinges on the set of templates chosen and whether the limited basis spanned by these templates contains a good approximation for the true properties of these galaxies. This is a particularly strong assumption when observing in a new regime since the template basis must be chosen based on extrapolations from existing observations.

Fortunately, follow-up spectroscopy has typically shown that photometric redshifts have generally been a good predictor of spectroscopic redshifts \citep{Hildebrandt_2010}. It has thus been assumed that the properties produced by template fitting are similarly good predictors of true galaxy properties, although there is less direct evidence of these properties to use as a basis for comparison.

However, with the Cycle 1 observations from the James Webb Space Telescope (JWST), several candidates for $z>15$ galaxies have been reported using photometric template fitting \citep{10.1093/mnras/stac3472, Castellano_2022, Yan_2022}, but spectroscopic follow-ups and photometric observations at other wavelengths have shown that these galaxies are much lower-redshift objects \citep{ArrabalHaro2023, Zavala_2023}. Additionally, many $z\sim10$ candidates have indeed been found to lie at high redshift but had spectroscopic redshifts $10\%$ to $20\%$ lower than those fitted photometrically \citep{Fujimoto_2023}. 
These performances can in part be attributed to spectral "dropouts" (e.g., photometric filters returning zero-values), which in effect limits information for higher redshift objects. However, these discrepancies also indicate that revised templates could be crucial to probing these early galaxies.

Several approaches have been taken to revise templates in order to better describe galaxies and their stellar populations at $z \sim 10$. Some template sets have been produced based on using astrophysical arguments to choose parameters in stellar population synthesis (SPS) models that should better reflect early galaxies \citep{Steinhardt_2023,Larson_2023,Adams_2022}. Alternatively, the approach chosen by Turner (in prep.) restricts SPS models to those with younger stellar populations. Finally, a hybrid approach consists of taking star formation histories from early universe simulations and synthesizing a stellar population by including an age restriction (see table \ref{tab:templates}, \citealt{Brammer_2008}).

\begin{table}[t]
\renewcommand{\arraystretch}{1.3}
\caption{Template sets used. \label{tab:templates}}
\centering
\small 
\begin{tabular}{lccc} 
\hline\hline
Name & Abbr. & Size \\
\hline
Blue-SFHz-13\tablefootmark{(1)}   & \texttt{BlSFH}\label{BlSFH} & 14\tablefootmark{(*)}  \\
            Carnall-SFHz-13\tablefootmark{(1)}   & \texttt{CaSFH}\label{CaSFH} & 14\tablefootmark{(*)}  \\
            EAZY-v1.3\tablefootmark{(1)}  & \texttt{Ev3}\label{EAZYv13} & 9  \\
            EMLines\tablefootmark{(2)}   & \texttt{A22}\label{EMLines} & 39  \\
            Fsps-45K\tablefootmark{(3)}   & \texttt{S23}\label{FSPS45K} & 6  \\
            Larson-SED-template\tablefootmark{(4)}   & \texttt{L23}\label{Lar23} &  18\\ 
            fsps-v3.2-Chabrier03\tablefootmark{(5)}\tablefootmark{(**)}  &  \texttt{T22}\label{Wil22} & 12 \\
            tweak-fsps-QSF-12-v3\tablefootmark{(1)}   & \texttt{TwSPS}\label{TwSPS} & 12\\
\hline
\end{tabular}
\tablefoot{
The eight template sets used here were selected to span the range of current models. The template sets \texttt{BlSFH}, \texttt{CaSFH}, \texttt{Tweak\_fsps}, and \texttt{Ev3} were included with the current version of the EAZY software package. 
\tablefoottext{*}{The given size is the number of templates in the full basis; however, the SFHz templates employ a variable template size dependent on redshift. With fewer templates available at the highest redshifts, this makes them effectively more constrained.} \tablefoottext{**}{The author provides a large set of templates for use. In preliminary tests, their redshift performance was found to be largely homogeneous; however, the one listed above was found to perform marginally better, and so was selected for use.}
The template sets used here were taken from: \tablefoottext{1}{\citealt{Brammer_2008};} \tablefoottext{2}{\citealt{Adams_2022};} \tablefoottext{3}{\citealt{Steinhardt_2023};} \tablefoottext{4}{\citealt{Larson_2023};} \tablefoottext{5}{Turner (in prep.)}
}
\end{table}

This variety of meaningful approaches means that astronomers are now forced to choose which template to use. In this work, different template sets and techniques are evaluated in order to determine which is the most robust. Currently, photometric template fits are used in two different ways. In some cases, only the photometric redshift is relevant since high-redshift candidates are then targeted for follow-up spectroscopy in order to better determine their properties. In other cases, physical parameters are inferred directly from photometric template fits and used to draw scientific conclusions without additional observations. As a result, two different tests are conducted, one to determine which template basis is best at redshift estimation and the other to determine which is most successful at recreating spectral features.

In Sect. \ref{sec:samples}, the various datasets available for photometry and spectroscopy with high-redshift sources (here defined as $z>5$) are described along with the template sets used to fit them. The fitting technique is then described in Sect. \ref{sec:methodology}. In Sect. \ref{sec:results}, the template sets are evaluated on their ability to both provide photometric redshifts and predict the full SED. Finally, the implications for future surveys are discussed in Sect. \ref{sec:discussion}.
\section{Sample description}
\label{sec:samples} 
Any evaluation of template performance requires a benchmark dataset with robust measurements of properties that template sets can then be asked to predict. This requires spectroscopic observations in addition to the photometry, which will be made available for template fitting.  Here, the spectroscopy was drawn from several JWST catalogs, with photometry supplemented by any additional available observations.

The template sets were evaluated based on both redshift prediction and SED feature prediction since a template might excel in one and not the other. This evaluation was thus done with two separate tests and two separate samples.

An SED sample was chosen to determine whether reconstructed SEDs correctly predict observed SEDs. Many of the differences between models lie in the strength of the emission lines, which have been observed to be unexpectedly strong at higher redshifts \citep{10.1093/mnras/stae2430,Trump_2023}. Therefore, this test required sufficiently deep spectra to separate the continuum emission from the spectral lines and to measure the continuum shape. This was further complicated by the significant discrepancies in spectral properties from independent spectroscopic catalogs \citep{Mugnai_2024} reporting the same JWST observations. This is likely due to differing assumptions about properties such as slit loss. Therefore, the SED sample was highly restricted to only the objects with clean and simple reduction. The result is a limited but robust test, where negligible future revisions are needed as spectroscopic reductions improve.

A far larger  redshift sample was also chosen in order to determine which template sets produced photometric redshifts ($z_{phot}$) that most closely mirror spectroscopic redshifts ($z_{spec}$). This spectroscopic sample is more robust since it is only necessary to demonstrate a few emission lines rather than a continuum. In addition, although independent spectroscopic catalogs often report differences in the strength of emission lines, there is a broad agreement on the observed wavelength and thus the redshifts. As a result, the redshift sample is far larger, allowing comprehensive results and subsampling.

\subsection{Redshift sample}
\label{sec:samples_}
\begin{figure}[ht]
    \centering
    \includegraphics[width=0.49\textwidth]{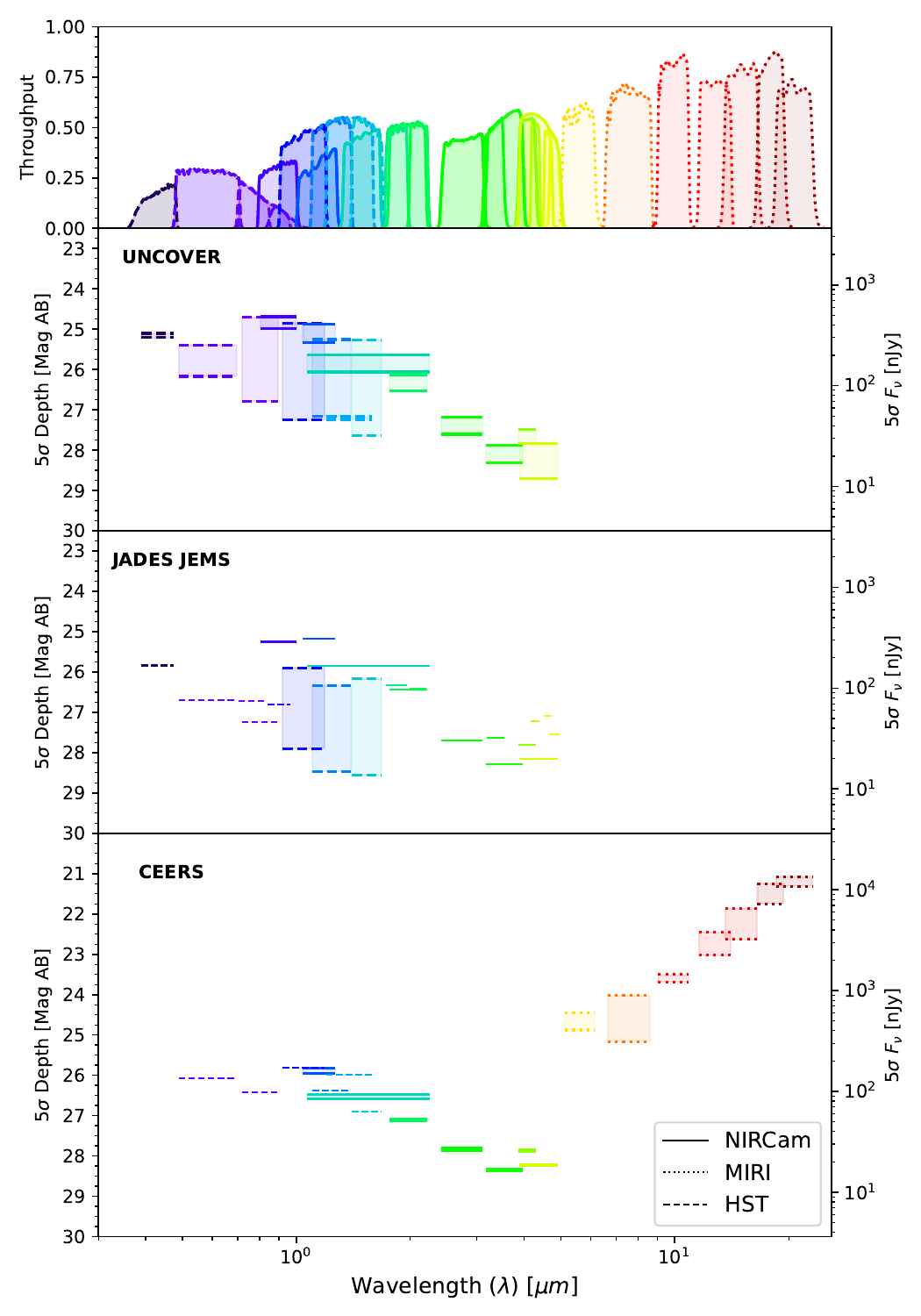}
    \caption{\label{fig:filtercoverage}  Filters covered in the surveys used and the reported observational depth of individual filters. Solid lines ("---") indicate JWST NIRCam filters, dashed lines \mbox{("- -")} indicate filters from the HST, and dotted lines ("$\cdots$") indicate filters from the JWST MIRI instrument. Apertures were set variably according to the Rayleigh criterion formula and thus range from $0.035^{\prime\prime}$ to $0.81^{\prime\prime}$. The depth data presented was derived from reported values by survey authors \citep{Weaver_2023,deugenio2024jadesdatarelease3,Rieke_2023,Hainline_2024,Guo_2013,Bagley_2023,Yang_2023}. Some surveys employ in-house processes to the Hubble data and thus report adjusted depths.}
\end{figure}

The redshift sample was used as a benchmark sample to evaluate the accuracy of the photometric redshift. The benchmark sample consists primarily of  $z_{spec}>5$ objects, which shall henceforth be defined to be high-redshift. However, with practical analysis of high-redshift sources being inseparable to knowing a prior of whether an object is high-redshift in the first place, evaluation of an all-regime sample was also necessary (here cut by $z_{spec}>0.1$). Datasets from three surveys were used for this analysis, collecting the following object counts:

 114 objects in the high-redshift sample from CEERS: There are 76\,637 total objects in The Cosmic Evolution Early Release Science Survey \citep[CEERS; August 30, 2023 - update used;][]{CEERS_2023, Bagley_2023, Yang_2021}, in a combined dataset with data from the Hubble legacy field (HLF). CEERS employs photometric exposure times between $\SI{1.5}{ks}$ and $\SI{3.1}{ks}$, achieving $5\sigma$ per-filter depths between $\SI{26.67}{mag_{AB}}$ and $\SI{29.22}{mag_{AB}}$ (avg. $\SI{28.4}{mag_{AB}}$), reported with $0.2^{\prime\prime}$ diameter aperture. Photometry includes bands in Near Infrared (NIR) and Mid Infrared (MIR), with additional bands from the HLF, all with varying coverage, as summarized in Figure \ref{fig:filtercoverage}. A combined photometric catalog was provided by the DAWN JWST Archive (DJA).\footnote{\url{https://dawn-cph.github.io/dja/}}

Spectroscopy \citep{Arrabal_Haro_2023} is obtained in parallel using both the micro-shutter assembly (MSA) and wide field slitless spectroscopy (WFSS), providing many robust $z_{spec}$ measurements. A total of 466 objects have $z_{spec}>0.1$ and were thus included in the all-range redshift sample, and 114 of these have $z_{spec}>5$ placing them in the high-redshift sample. The survey, generally having relatively shorter exposures (see Figure \ref{fig:filtercoverage}) but broad wavelength coverage, yields a large and relatively shallow wide-field survey of high-redshift objects.

30 objects in the high-redshift sample from UNCOVER:  There are 61\,648 objects in the Ultradeep NIRSpec and NIRCam ObserVations before the Epoch of Reionization (UNCOVER) catalog \citep[downloaded 2024 March 6;][]{Bezanson_2022,Weaver_2023,Furtak_2023, Fujimoto_2023, Wang_2023} in a combined dataset from the Hubble Frontier Field (HFF; \citealt{Lotz_2017}), and other programs not utilized here. NIRCam measurements have exposures ranging from $\SI{9.4}{ks}$ to $\SI{21.6}{ks}$ (average ${\sim}\SI{15}{ks}$), yielding per-filter $5\sigma$ depths of ${\sim}\SI{29.5}{mag_{AB}}$ with variable aperture ($0.16^{\prime\prime}{-} 0.32^{\prime\prime}$). Bands were included from NIRCam with additional bands from the HFF, all summarized in Figure \ref{fig:filtercoverage}. The photometric catalog was provided by DJA. Crucially, the survey predominantly covers lensed regions, with expected magnifications between mild values of ${\sim}1.2$ and very high values of ${\sim}10$ \citep{Bezanson_2022}. This yields effective depths up to ${\sim}\SI{32}{mag_{AB}}$, albeit highly variable within the field.

Spectroscopy in the UNCOVER field includes both prior HST coverage and JWST's MSA and WFSS observations. This provides robust redshift measurements of 447 objects ($z_{spec}>0.1$), and 30 with $z_{spec}>5$ (values were collected from the DJA archive).

99 objects in the high-redshift sample from JADES: The JADES catalog contains 94\,000 objects in a combined catalog including JWST Advanced Deep Extragalactic Survey (JADES; \citealt{bunker2024jadesnirspecinitialdata,eisenstein2023overviewjwstadvanceddeep,Hainline_2024,Rieke_2023,eisenstein2023jadesoriginsfieldnew,deugenio2024jadesdatarelease3}), JWST Extragalactic Medium-band Survey \citep[JEMS;][]{Williams_2023} and the Hubble Legacy Field (HLF) with parts of the Hubble Ultra Deep Field (HUDF; downloaded on 2024 March 6). JADES has mixed exposure times between $\SI{8.7}{ks}$ and $\SI{77.6}{ks}$, achieving $10\sigma$ depths of ${\sim}\SI{29.7}{mag_{AB}}$ with $0.2^{\prime\prime}$ aperture diameter with NIR bands. The adjacent JEMS survey has exposure times of ${\sim}\SI{23.6}{ks}$, achieving $5\sigma$ depth of $\SI{23.6}{ks}$ with $0.3^{\prime\prime}$ aperture diameter in MIR bands. The HLF adds additional coverage in optical bands. A summary of band coverage is provided in Figure \ref{fig:filtercoverage}.

Spectroscopy is measured through follow-up observations using JWST's MSA and from the HST, achieving reliable redshifts for 446 objects with $z_{spec}>0.1$ (predominantly from the DJA archive; \citealt{brammer_2023_8319596,doi:10.1126/science.adj0343}). 99 objects have $z_{spec}>5$ and were thus included in the high-redshift sample. 

JADES includes variable coverage across the survey rather than a fixed number of available bands (e.g., there are 74 objects with 20 bands available). In brief, JADES is a prototypical example of a large "conventional" (non-lensed) JWST survey with deep measurements in NIR bands and additional shallow measurements in MIR bands. Furthermore, the inclusion of MSA metadata allowed the use of several spectra in the SED sample, as described below.

In total, from these three surveys, $143$ objects were included in the high-redshift sample, and $1359$ objects in the all-regime sample. In the all-regime sample, 466 objects were included from CEERS, 447 from UNCOVER, and 446 from JADES.

\subsection{SED subsample}
\label{sec:samples_sed}

\begin{figure}[ht]
    \centering
    \includegraphics[width=0.49\textwidth]{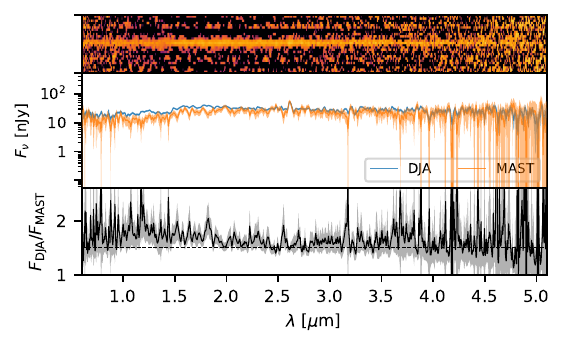}
    \caption{\label{fig:discrepancy}   Example of reduction discrepancy between MAST and DJA reductions of JADES ID = 4413 (MAST NIRSpec ID). The object was selected arbitrarily as an example with typical rather than maximal reduction discrepancy. A stacked combination of available dispersers is shown (middle plot), along with the raw 2D Prism slit measurements (logarithmic color-scale, $F_\nu$, top plot).  The ratio $F_{\nu,DJA}/F_{\nu,MAST}$ is also color-dependent (bottom plot). The SEDs are sampled at R = 1000. Noteworthy discrepancies include the overall uncertainty difference, and a blue tendency in the DJA reduction as compared to MAST. Furthermore, feature differences around the $1{-}2\mu m$  and $3{-}4\mu m$ ranges are present. The reductions used here were collected from: \citealt{deugenio2024jadesdatarelease3,Brammer_2023}.}
\end{figure}

The high-redshift SED subsample required spectroscopy sufficiently deep and resolved to be able to measure the continuum emission in addition to spectral lines. Specifically, the SED subsample was restricted to spectra from the high-redshift sample with a continuum average $\text{signal-to-noise ratio (S/N)}\geq5$, and $R>1000$ (where $R$ is the spectral resolution, defined as $R=\frac{\lambda}{\Delta\lambda}$, with $\Delta\lambda$ being the smallest distinguishable wavelength at a given wavelength).

In addition, due to significant discrepancies between the reduced versions of the same spectra from different catalogs, further restrictions were applied to the SED subsample to minimize these effects (see Fig. \ref{fig:discrepancy}). These discrepancies are likely due to complicated distortions from neighboring objects, slit misalignment effects, assumptions about the shape of the source, and differing techniques employed to mitigate those differences. To decrease contamination from neighboring sources, only slit spectroscopy was used (specifically MSA). Additionally, objects were constrained to not have neighboring emission within the slit length of ${\sim}0.46^{\prime\prime}$ (thus angular distance added to the Point Spread Function (PSF) Full Width at Half Maximum (FWHM) radius must be ${\geq}0.46^{\prime\prime}$). To further simplify the data reduction of a given object, a selection was performed toward "centered" objects, where the majority of the object's flux must be enclosed in the slit. This minimized the influence of potentially dissimilar methods for accounting for slit-diffraction effects. In practice, this limited the dataset to a subset of JADES where both dimensions of slit-offset were available in the metadata. Finally, sources were limited to only point-like sources (FWHM$\leq0.15^{\prime\prime}$).

\begin{figure}[ht]
    \centering
    \includegraphics[width=1\linewidth]{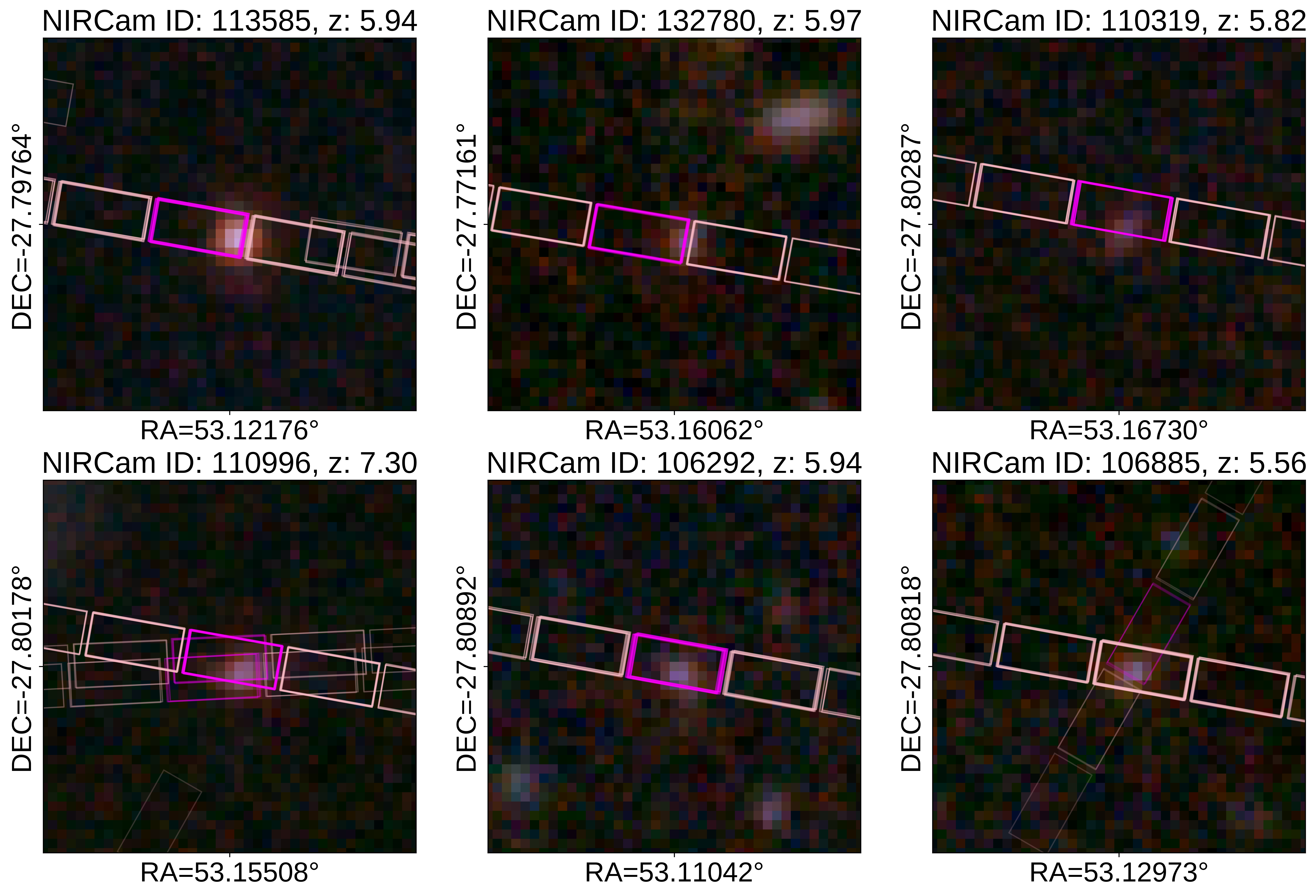}
    \caption{ Objects in the SED subsample. Selected by the criteria above from JADES Data Release 2. Cutouts are selected from the DJA archive and are produced with the GRIZLI utility \citep{Brammer2023-wi}. Redshifts, catalog NIRCam ID, and coordinates are noted. F150W, F277W, and F444W was used to make an RGB cutout with equal weighing. The MSA shutter placement is overlaid in orange, with primary source slits marked in magenta.}
    \label{fig:specobjs}
\end{figure}

In summary, the additional cuts made to the SED subsample with the high-redshift sources were as follows:
\begin{enumerate}
    \item Combined-grating average $\text{S/N}\geq 5$ in the rest frame $\SI{1000}{\angstrom}$ to $\SI{5000}{\angstrom}$ with $R=1000$: This signal strength requirement was necessary in order to meaningfully compare the distribution between the continuum and the line flux from the best-fit and observed SEDs. Eddington bias was introduced as a consequence and should be considered when evaluating the results.
    \item R $\geq$ 1000: The sample was limited to objects with highly resolved SED measurements to ensure resolved spectral features for the comparison. If available, additional measurements with $R < 1000$ of the same object (for instance, a measurement using the JWST PRISM disperser) were combined to observations with $R > 1000$, upon the condition that a majority of the SED had spectroscopic coverage with $R > 1000$.
    \item Only MSA spectroscopy with $R_{FWHM}/2$ within the slit opening: This ensured the majority of the object flux was within the slit, safeguarding against anomalies in the compensation of slit-edge effects. It also put an upper bound on object radius, effectively limiting sources to point-like sources. This decouples spectra quality from any assumptions of object shape or size in the reduction, while biasing toward lower observed object sizes. This yields a heavy bias toward both object type and distance.
    \item No other object with its $R_{FWHM}$ $F444W$ emission within $0.46^{\prime\prime}$ of target object ($0.46^{\prime\prime}$ being the length of the micro-shutters): This was done to prevent other objects from contaminating observations within the same slit. This cut eliminates known cases of two objects located within one slit, but at the same time it removes any mergers or doubly cataloged galaxies.
\end{enumerate}

As with any cuts, these choices introduced biases into the sample, some of which may be significant. This included an Eddington bias with the continuum S/N limit, and a bias toward higher redshift or smaller objects with the point-like source requirement. Perhaps the strongest bias was that the selection was biased against objects such as disk-like galaxies, with higher angular momentum, mass, or both, since these typically have a larger source diameter \citep{Romanowsky_2012}. However, these quality cuts were deemed necessary in order to produce a robust sample.

JADES is the only survey with the required metadata to perform these cuts easily available. The survey measures its spectra with the JWST parallel (MSA). It consists of two deep pointings of $\SI{192}{ks}$ exposure, 14 "medium-depth" pointings of ${\sim}\SI{43}{ks}$, and a shallow set of 15 with ${\sim}\SI{14}{ks}$ exposure times. The observations of each object include a coarse prism observation with $R\sim30{-}300$, three medium dispersion $R\approx1000$ gratings, and one $R\approx2700$ grating, giving a combined overlapping coverage over $\SI{0.6}{}{-}\SI{5.3}{\mu m}$.

These quality restrictions yielded six objects for use, summarized in Figure \ref{fig:specobjs}. Given the small sample size, it is likely that the SED sample is not fully representative of the high-redshift galaxy population. However, the sample is still sufficient to draw conclusions of ill-fit sources.

\subsection{Template sets}
\label{sec:templSumm}
The templates included were chosen as a representative set of high-redshift templates in current use. Each of these templates have in their development somewhat started with lower redshift templates, but made modifying choices designed for high-redshift application. Three broad approaches has been taken in choosing which modifications were needed for the high-redshift regime: (1) changes driven by empirical differences between low-redshift and high-redshift SEDs; (2) changes motivated by pen-and-paper theory, predicting the ways that high-redshift galaxies are predicted to be distinct; and (3) changes driven by simulations modeling the ways that high-redshift galaxies are predicted to be distinct.

Perhaps the most straightforward approach is to modify previous templates based on empirical differences between high-redshift galaxies and those at lower redshift.  This can both take the form of empirically motivated adjustment, or direct empirical changes. For example, $z \gtrsim 8$ galaxies have significantly stronger emission lines than previously predicted \citep{10.1093/mnras/stae2430,Trump_2023}. Thus, templates can be modified by re-scaling emission line strengths in order to better match observations, such as in the \texttt{A22} template \citep[see our Table \ref{tab:templates} and][]{Adams_2022}. Other similar approaches include \texttt{BlSFH} and \texttt{L23} \citep{Brammer_2008, Larson_2023}, motivated by favoring bluer spectra than predecessors; and \texttt{CaSFH} and \texttt{T22} (\citealt{Brammer_2008}; Turner in prep.) based on wider varieties of synthetic stellar populations with varying properties. A significant disadvantage of a purely empirical approach is that the templates are not necessarily produced by underlying astrophysical models, making it more difficult to infer astrophysical parameters from the best-fit spectrum. Thus, a model-driven approach would be preferred if it can reproduce observed SEDs equivalently well.

An alternate approach is to use theoretical predictions about the properties of the first galaxies to modify astrophysical parameters, then to use stellar population synthesis codes to turn those parameters into template sets. For example, because the cosmic microwave background temperature at $z > 7$ exceeds temperatures in galactic star-forming regions, the IMF is expected to be bottom-lighter than previously assumed \citep{10.1093/mnras/sty2123,Sneppen_2022,Steinhardt_2023}. Similarly, an unconstrained fit often produces stellar populations or other features of the SED that must be older than the presumed age of the universe under the standard $\Lambda$CDM cosmological paradigm \citep{Steinhardt_2024}. Thus, limited template sets have been constructed that are constrained to remain consistent with $\Lambda$CDM (\texttt{BlSFH}, \texttt{CaSFH}, \texttt{S23}; \citealt{Brammer_2008, Steinhardt_2023}). \texttt{T22} has taken a similar approach, adjusting population synthesis parameters to better represent high-redshift populations, with a "bursty" SFH and generally younger stellar populations (Turner 2024 priv. comm.).

Another approach is to derive spectra from the properties of the first galaxies from numerical simulations. Here, template sets are intended to be representative of the properties of a wide array of simulated galaxies, while avoiding the inclusion of properties that are uncharacteristic of simulated results. This approach has been taken by \citep{Larson_2023}, with the goal of producing bluer templates to better match observations at high-redshift. These bluer spectra are added to a previous template "tweak\_fsps\_QSF\_v12\_v3" \citep[\texttt{TwSPS};][]{Brammer_2008}. \texttt{TwSPS} was also added as a control sample since \texttt{L23} is an extension to this template set.

Finally, a template originally developed for lower redshift, "eazy\_v1.3" \citep{Brammer_2008}, was included as a control sample. No modifications have been made to tune this template set to the properties of high-redshift galaxies. Thus, it is expected that if the differences are significant, this lower redshift template set should perform relatively poorly.

Determining which of these approaches are the most successful at reconstructing observed SEDs, is also a tracer for gauging the current state of template fitting and the underlying physical models. If the empirical approaches are better, it would suggest that the current models of the first galaxies are still insufficient at inferring meaningful astrophysical parameters. In this case, only redshifts and properties that can be calculated from the redshift alone, such as luminosity functions, would be meaningful. However, if the theoretical approach is more successful, it would indicate that the major differences between the first galaxies and subsequent evolution are now well understood. Similarly, if simulation-driven approaches are successful, it would indicate that the constrained set of astrophysics included is adequate for the high-redshift regime.

\section{Methodology and metrics}
\label{sec:methodology}
As described in Sect. \ref{sec:samples}, two tests were performed, one of redshift accuracy for the 243-object high-redshift sample and 1359-object all-range sample, and another of full SED reconstruction for the 6-object SED sample. All template fits were done with EAZY, as it provides a platform to exhaustively explore the entire template fit search space up to a few minor nonlinear fit improvements (G. Brammer, private communication).

By default, EAZY uses a luminosity prior function when assessing highest likelihood redshifts. It should be noted that some results might be sensitive the choice of prior (see \citealt{Brammer_2008} for the specific prior distribution). Here, the default recommended EAZY settings were chosen since those were likely to be the settings most commonly used, and a proper investigation of alternative priors was left to future work.

\subsection{Redshift performance}
\label{sec:redshiftperformance}
The redshift test followed the methodology outlined in \citet{Hildebrandt_2010}, in order to use established metrics for testing photometric redshift performance. All comparisons are expressed in terms of the fractional difference between the photometric and spectroscopic redshifts,
\label{equ:deltaz}
\begin{equation}
    \Delta z \equiv \frac{(1 + z_{phot}) - (1 + z_{spec})}{1+z_{spec}} = \frac{z_{phot}-z_{spec}}{1+z_{spec}}.
\end{equation}

Template sets were then compared on three initial criteria, each potentially most important under different circumstances. For the purposes of producing a sample suitable for follow-up spectroscopy, the most important metric is the catastrophic error fraction $\eta\equiv \frac{N\left(\left|\Delta z\right|>0.15\right)}{N}$. However, often if photometric redshifts are to be used directly, the scatter $\sigma_{nmad}\equiv \sigma\left(\Delta z\right)$ or bias $\Delta_{bias}\equiv \overline{\Delta z}$ might be more important. 

One may naturally expect the set of metrics \{$\Delta_{bias}$, $\sigma_{nmad}$, $\eta$\} to be well correlated, in that a very poor fitting procedure is likely to produce photometric redshifts that perform poorly in all three metrics. Similarly, if a model better describes the underlying astrophysics, it would naturally perform better in all three. However, it is also possible to over-tune a model to a limited dataset such as early JWST observations. In such cases, there will instead be a trade-off between bias and variance \citep{vonluxburg2008statisticallearningtheorymodels,10.1162/neco.1992.4.1.1}. As models become more complex and use additional parameters to match a training set, the bias decreases. Since additional tuning may not describe the data better, the model will typically produce a higher variance when fit to a new sample. Specific science results might depend almost solely on the bias or variance alone, but for more purposes, a balance of low bias and low variance is optimal.

There is a significant variation between the different catalogs for the same JWST photometry. Therefore, in estimating uncertainties, it was assumed that a significant source of error might be attributed to unspecified systematics within the encompassing JWST photometry pipeline. Estimated errors were assigned to each photometric band and propagated using bootstrapping within the attained errors for each band. Errors were conservatively set to a 2\% baseline, to both account for the ${<}1\%$ reported calibration and potential reduction systematics. F277W, F430M, F460M, and F470M, being filters with reported greater systematics, were cautiously assigned 6\%, 4\%, 4\%, 8\%, respectively given reported pipeline issues \citep{STScI_NIRCam_Imagin_Calibration_Status_-_JWST_User_Documentation_2024}. These errors are propagated through the fitting procedure to attain distributions of redshift metrics. Average values are reported with $3\sigma$ deviations. Combined with the conservative bounds, this ensures that the reported metrics are robust toward even major future systematics mitigation. ${\geq}250$ resamples were produced until convergence of the metrics (specifically ${\sim}250$, ${\sim}250$, ${\sim}350$ for UNCOVER, CEERS, JADES respectively).

\subsection{SED performance}
The SED performance was tested by how well the photometrically reconstructed SEDs from templates reproduced the measured SEDs. In the reconstruction, EAZY might have found incorrect redshifts to be more favorable for reconstructing the SED. To standardize the test, the reconstructed SEDs were specifically chosen to be the best matches at $z_{spec}$. Here, $\chi^2_{\nu}$ was compared against the observed SED across templates to give a quantitative measure of template performance. 

The differences between the observed and reconstructed SEDs comes from two primary sources: (1) spectral features that are unrepresented in template sets; and (2) a lack of information available from the photometric bands with sufficient S/N, as opposed to a higher-resolution spectrum. To determine which was more significant, a secondary fit was produced with information from spectroscopic observations calibrated to photometry, where again the best fit at $z_{spec}$ was obtained. This enabled a more controlled analysis of whether a template represents an object well, as opposed to the photometry-based fit, where other effects such as lacking information, degeneracies or strong covariances can affect the fit.

\section{Template performance}
\label{sec:results}

Using the samples described in Sect. \ref{sec:samples} and methods described in Sect. \ref{sec:methodology}, all eight template sets (Table \ref{tab:templates}) are evaluated based on their predictive performance in the redshift and SED samples. The results are as follows.

\subsection{Redshift analysis}
\label{zanalysis}

\begin{table*}[ht]
\renewcommand{\arraystretch}{1.3}
\caption{\centering JADES $z>5$ photometric redshift performance metrics.}
\label{tab_zs_stats_JADES_high}
\centering
\small 
\begin{tabular}{c|lllll}
\hline\hline
Template & \thead{$\Delta_{\mathrm{bias},z>5}$} & \thead{$\sigma_{\mathrm{nmad},z>5}$} & \thead{$\eta_{f,z>5}$} & \thead{$\eta_{c,z>5}$} & \thead{$s_{z>5}$}\vspace{-0.1cm} \\ 
 & \thead{$(\%)$} & \thead{$(\%)$} & \thead{$(\%)$} & \thead{$(\%)$} & \thead{$(\%)$}\\
\hline
\texttt{A22} & $-3.7\pm0.2$ & $19.3\pm0.3$ & $12.0\pm0.3$ & $13.3^{+0.3}_{-0.4}$ & $98.74\pm0.07$ \\
\texttt{BlSFH} & $-3.4\pm0.2$ & $18.3\pm0.3$ & $9.2\pm0.3$ & $9.4\pm0.3$ & $99.82^{+0.08}_{-0.11}$ \\
\texttt{CaSFH} & $-4.1\pm0.2$ & $19.4\pm0.4$ & $10.6\pm0.4$ & $11.8\pm0.4$ & $98.8\pm0.1$ \\
\texttt{Ev3} & $-4.5\pm0.3$ & $22.4\pm0.4$ & $14.7\pm0.4$ & $15.2\pm0.4$ & $99.5\pm0.1$ \\
\texttt{L23} & $-2.2\pm0.2$ & $17.9\pm0.4$ & $8.8\pm0.3$ & $9.1\pm0.3$ & $99.72^{+0.08}_{-0.09}$ \\
\texttt{S23} & $-1.6\pm0.2$ & $16.0\pm0.3$ & $9.8\pm0.3$ & $9.9\pm0.3$ & $99.88\pm0.05$ \\
\texttt{T22} & $-1.7\pm0.3$ & $17.6\pm0.4$ & $9.7\pm0.4$ & $9.8\pm0.4$ & $99.82\pm0.08$ \\
\texttt{TwSPS} & $-8.4\pm0.5$ & $26.8\pm0.5$ & $15.8\pm0.6$ & $24.8\pm0.9$ & $90.9^{+0.6}_{-0.7}$ \\
\hline
\end{tabular}
\tablefoot{Overview of template performance within the $z_{spec}>5$ JADES sample. Specific metrics are defined above (Sect. \ref{sec:redshiftperformance}). The JADES survey is typical of the expected performance of a conventional deep photometric survey in the high redshift regime. Initial observations include the portion of non-catastrophic estimations, where only \texttt{BlSFH} and \texttt{L22} get ${>}90\%$ with a complete sample. Further, a lower bias and deviation indicate \texttt{S23}, \texttt{T22}, and \texttt{L22} to be somewhat more indicative of underlying physics.}
\end{table*}

\begin{table*}
\renewcommand{\arraystretch}{1.3}
\caption{\centering UNCOVER $z>5$ photometric redshift performance metrics.}
\label{tab_zs_stats_UNCOVER_high}
\centering
\small \begin{tabular}{c|lllll} \hline\hline
\vspace{0.1pt}
Template & \thead{$\Delta_{\mathrm{bias},z>5}$} & \thead{$\sigma_{\mathrm{nmad},z>5}$} & \thead{$\eta_{f,z>5}$} & \thead{$\eta_{c,z>5}$} & \thead{$s_{z>5}$}\vspace{-0.1cm} \\ 
 & \thead{$(\%)$} & \thead{$(\%)$} & \thead{$(\%)$} & \thead{$(\%)$} & \thead{$(\%)$}\\
\hline
\texttt{A22} & $-5.1^{+0.4}_{-0.5}$ & $23.6\pm0.5$ & $9.7\pm0.5$ & $9.9\pm0.5$ & $99.8^{+0.1}_{-0.2}$ \\
\texttt{BlSFH} & $-5.2\pm0.6$ & $21.0^{+0.9}_{-1.0}$ & $7.7\pm0.7$ & $9.5\pm0.6$ & $98.2\pm0.4$ \\
\texttt{CaSFH} & $-6.5\pm0.6$ & $22.8\pm0.9$ & $9.8\pm0.9$ & $11.6^{+0.9}_{-0.8}$ & $98.2\pm0.4$ \\
\texttt{Ev3} & $-21.8\pm0.7$ & $35.2\pm0.3$ & $32\pm1$ & $34.7\pm0.9$ & $97.0^{+0.4}_{-0.5}$ \\
\texttt{L23} & $-2.1\pm0.6$ & $17\pm1$ & $5.8^{+0.8}_{-0.7}$ & $7.8\pm0.6$ & $98.0\pm0.4$ \\
\texttt{S23} & $-1.7^{+0.2}_{-0.3}$ & $13.2\pm0.4$ & $5.0\pm0.3$ & $5.0\pm0.3$ & $100\pm0$ \\
\texttt{T22} & $-4.0\pm0.6$ & $23.4\pm0.7$ & $9.4\pm0.6$ & $9.7\pm0.6$ & $99.6\pm0.2$ \\
\texttt{TwSPS} & $-38\pm1$ & $39.7^{+0.2}_{-0.3}$ & $45\pm1$ & $53\pm1$ & $92\pm1$ \\
\hline
\end{tabular}
\tablefoot{Overview of template performance within the $z_{spec}>5$ UNCOVER sample. Specific metrics are defined above (Sect. \ref{sec:redshiftperformance}). The UNCOVER survey demonstrates the possibility of strong performance with the high depth of a lensed field despite fewer available filters. In such a survey, \texttt{S23} can achieve a complete sample with ${<}5\%$ catastrophic outliers.  However, there is a lower fit completeness rate across several templates, indicating missing representation. With both low bias and deviation, results suggest that \texttt{S23} is more physically representative.}
\end{table*}

\begin{table*}[t]
\renewcommand{\arraystretch}{1.3}
\caption{\centering CEERS $z>5$ photometric redshift performance metrics.}
\label{tab_zs_stats_CEERS_high}
\centering
\small 
\begin{tabular}{c|lllll} 
\hline\hline

Template & \thead{$\Delta_{\mathrm{bias},z>5}$} & \thead{$\sigma_{\mathrm{nmad},z>5}$} & \thead{$\eta_{f,z>5}$} & \thead{$\eta_{c,z>5}$} & \thead{$s_{z>5}$}\vspace{-0.1cm} \\ 
 & \thead{$(\%)$} & \thead{$(\%)$} & \thead{$(\%)$} & \thead{$(\%)$} & \thead{$(\%)$}\\
\hline
\texttt{A22} & $-5.5\pm0.6$ & $29.0^{+0.5}_{-0.4}$ & $26.1\pm0.6$ & $26.6\pm0.6$ & $99.5\pm0.1$ \\
\texttt{BlSFH} & $-11.9\pm0.5$ & $33.0\pm0.5$ & $27.7\pm0.7$ & $28.4\pm0.7$ & $99.3\pm0.1$ \\
\texttt{CaSFH} & $-16.4\pm0.5$ & $34.9^{+0.4}_{-0.5}$ & $35.8\pm0.7$ & $36.7\pm0.7$ & $99.1\pm0.1$ \\
\texttt{Ev3} & $-10.0^{+0.5}_{-0.6}$ & $33.6\pm0.5$ & $32.0^{+1.0}_{-0.9}$ & $33.5\pm0.9$ & $98.5\pm0.2$ \\
\texttt{L23} & $-7.2\pm0.7$ & $30.9\pm0.6$ & $26.3\pm0.7$ & $27.3\pm0.7$ & $98.94^{+0.07}_{-0.08}$ \\
\texttt{S23} & $-2.2\pm0.3$ & $22.6\pm0.5$ & $22.4\pm0.7$ & $22.4\pm0.7$ & $99.996^{+0.004}_{-0.014}$ \\
\texttt{T22} & $-2.9\pm0.5$ & $28.2\pm0.5$ & $24.1\pm0.7$ & $25.3\pm0.7$ & $98.8\pm0.1$ \\
\texttt{TwSPS} & $-30\pm1$ & $43.4\pm0.4$ & $46.3\pm0.8$ & $56\pm1$ & $90.0\pm0.6$ \\
\hline
\end{tabular}
\tablefoot{Overview of template performance within the $z_{spec}>5$ CEERS sample. Specific metrics are defined above (Sect. \ref{sec:redshiftperformance}). The CEERS survey has a lower depth, and it can be seen that this requires caution when fitting in the high-redshift regime. In bias, deviation, and catastrophic fit proportion, \texttt{S23} performs best, perhaps due to being supported by its smaller basis and thus larger constraint.}
\end{table*}

\begin{table*}[ht]
\renewcommand{\arraystretch}{1.3}
\caption{\centering JADES $z>0.1$ photometric redshift performance metrics.}
\label{tab_zs_stats_JADES_all}
\centering
\small
\begin{tabular}{c|lllll} 
\hline\hline
\vspace{0.1pt}
Template & \thead{$\Delta_{\mathrm{bias},z>0.1}$} & \thead{$\sigma_{\mathrm{nmad},z>0.1}$} & \thead{$\eta_{f,z>0.1}$} & \thead{$\eta_{c,z>0.1}$} & \thead{$s_{z>0.1}$}\vspace{-0.1cm} \\ 
 & \thead{$(\%)$} & \thead{$(\%)$} & \thead{$(\%)$} & \thead{$(\%)$} & \thead{$(\%)$}\\
\hline
\texttt{A22} & $-2.3\pm0.2$ & $37.9^{+0.8}_{-0.7}$ & $21.0\pm0.3$ & $21.5\pm0.3$ & $99.48^{+0.04}_{-0.05}$ \\
\texttt{BlSFH} & $-3.7\pm0.3$ & $34.4^{+0.9}_{-0.8}$ & $18.7^{+0.5}_{-0.4}$ & $19.4^{+0.5}_{-0.4}$ & $99.29^{+0.06}_{-0.09}$ \\
\texttt{CaSFH} & $-4.9\pm0.3$ & $34.0^{+0.9}_{-0.8}$ & $21.5\pm0.5$ & $22.5\pm0.5$ & $98.97^{+0.07}_{-0.1}$ \\
\texttt{Ev3} & $-5.7\pm0.3$ & $39.6\pm0.8$ & $25.8\pm0.4$ & $26.9\pm0.4$ & $98.89^{+0.08}_{-0.09}$ \\
\texttt{L23} & $-4.7\pm0.3$ & $38.4\pm0.8$ & $23.5^{+0.5}_{-0.4}$ & $24.6\pm0.5$ & $98.9\pm0.1$ \\
\texttt{T22} & $-1.0^{+0.3}_{-0.4}$ & $39\pm1$ & $21.4^{+0.5}_{-0.4}$ & $21.9\pm0.5$ & $99.46\pm0.06$ \\
\texttt{TwSPS} & $-12.1\pm0.4$ & $42.9^{+1.0}_{-0.9}$ & $32.1\pm0.5$ & $36.6^{+0.6}_{-0.5}$ & $95.5\pm0.2$ \\
\hline
\end{tabular}
\tablefoot{Overview of template performance in the all-redshift ($z_{spec}>0.1$) JADES sample. 
Specific metrics are defined above (Sect. \ref{sec:redshiftperformance}). With the main purpose of the "all-regime" sample being the selection of a high-redshift sample, it is clear that \texttt{BlSFH} performs best. The sets \texttt{T22} and \texttt{S23} are not included in this analysis because they require pre-selection of high-redshift candidates.}
\end{table*}

\begin{table*}
\renewcommand{\arraystretch}{1.3}
\caption{\centering UNCOVER $z>0.1$ photometric redshift performance metrics.}
\label{tab_zs_stats_UNCOVER_all}
\centering
\small 
\begin{tabular}{c|lllll} 
\hline\hline
Template & \thead{$\Delta_{\mathrm{bias},z>0.1}$} & \thead{$\sigma_{\mathrm{nmad},z>0.1}$} & \thead{$\eta_{f,z>0.1}$} & \thead{$\eta_{c,z>0.1}$} & \thead{$s_{z>0.1}$}\vspace{-0.1cm} \\ 
 & \thead{$(\%)$} & \thead{$(\%)$} & \thead{$(\%)$} & \thead{$(\%)$} & \thead{$(\%)$}\\
\hline
\texttt{A22} & $-11.7\pm0.5$ & $28.0\pm0.4$ & $25\pm1$ & $25\pm1$ & $99.76^{+0.08}_{-0.12}$  \\
\texttt{BlSFH} & $-13.5^{+0.5}_{-0.6}$ & $27.1\pm0.3$ & $24.1^{+1.1}_{-0.9}$ & $25\pm1$ & $99.1^{+0.2}_{-0.3}$   \\
\texttt{CaSFH} & $-14.0^{+0.5}_{-0.6}$ & $26.9\pm0.3$ & $26\pm1$ & $28\pm1$ & $98.0^{+0.3}_{-0.4}$  \\
\texttt{Ev3} & $-18.6^{+0.5}_{-0.6}$ & $31.6\pm0.2$ & $32.6^{+0.9}_{-0.8}$ & $33.4^{+1.0}_{-0.9}$ & $99.2\pm0.2$  \\
\texttt{L23} & $-15.0^{+0.6}_{-0.7}$ & $28.7\pm0.3$ & $26.2^{+1.1}_{-0.9}$ & $27\pm1$ & $98.7^{+0.4}_{-0.5}$  \\
\texttt{T22} & $-10.3\pm0.5$ & $27.1\pm0.4$ & $22.1^{+1.0}_{-0.9}$ & $22.6^{+1.0}_{-0.9}$ & $99.5^{+0.1}_{-0.2}$  \\
\texttt{TwSPS} & $-33.0^{+0.8}_{-0.9}$ & $35.0\pm0.2$ & $48\pm1$ & $54\pm1$ & $94.6^{+0.5}_{-0.7}$  \\
\hline
\end{tabular}
\tablefoot{Overview of template performance on the all-redshift ($z_{spec}>0.1$) UNCOVER sample. Specific metrics are defined above (Sect. \ref{sec:redshiftperformance}). Within UNCOVER, sample selection of high-redshift galaxies seems largely independent of the templates as long as they are designed with high redshift in mind. With \texttt{BlSFH}, the fraction of successful photometric redshifts is ${\sim}75\%$. 
 Despite the increased depth due to lensing in UNCOVER, the overall performance of template fitting in the all-redshift sample is not improved over JADES (as opposed to the high-redshift sample).  This is most likely due to the more complicated foreground field associated with lensed surveys \citep{Vujeva_2024}. The sets \texttt{T22} and \texttt{S23} are not included in this analysis because they require pre-selection of high-redshift candidates.}
\end{table*}

\begin{table*}
\renewcommand{\arraystretch}{1.3}
\caption{\centering CEERS $z>0.1$ photometric redshift performance metrics.}
\label{tab_zs_stats_CEERS_all}
\centering
\small
\begin{tabular}{c|lllll}
\hline\hline

Template & \thead{$\Delta_{\mathrm{bias},z>0.1}$} & \thead{$\sigma_{\mathrm{nmad},z>0.1}$} & \thead{$\eta_{f,z>0.1}$} & \thead{$\eta_{c,z>0.1}$} & \thead{$s_{z>0.1}$}\vspace{-0.1cm} \\ 
 & \thead{$(\%)$} & \thead{$(\%)$} & \thead{$(\%)$} & \thead{$(\%)$} & \thead{$(\%)$}\\
\hline
\texttt{A22} & $-0.6\pm0.4$ & $50\pm1$ & $30.5^{+0.8}_{-0.7}$ & $31.9^{+0.8}_{-0.7}$ & $98.6\pm0.1$ \\
\texttt{BlSFH} & $-6.3\pm0.5$ & $45\pm1$ & $31\pm1$ & $32\pm1$ & $98.7\pm0.1$  \\
\texttt{CaSFH} & $-9.2\pm0.5$ & $45\pm1$ & $35.1^{+1.0}_{-0.9}$ & $37.8^{+1.0}_{-0.9}$ & $97.4\pm0.1$  \\
\texttt{Ev3} & $-6.1\pm0.5$ & $43.8\pm0.8$ & $32.4^{+1.0}_{-0.9}$ & $33.9^{+1.0}_{-0.9}$ & $98.5\pm0.1$ \\
\texttt{L23}& $-7.0\pm0.6$ & $51\pm1$ & $36\pm1$ & $38\pm1$ & $98.3\pm0.2$ \\
\texttt{T22}& $0.9^{+0.5}_{-0.6}$ & $50.1\pm0.9$ & $32\pm1$ & $33\pm1$ & $98.64^{+0.08}_{-0.09}$ \\
\texttt{TwSPS} & $-20.6\pm0.8$ & $54\pm1$ & $45\pm1$ & $52\pm1$ & $93.3^{+0.3}_{-0.4}$  \\
\hline
\end{tabular}
\tablefoot{Overview of template performance on the all-redshift ($z_{spec}>0.1$) CEERS sample.
Specific metrics are defined above (Sect. \ref{sec:redshiftperformance}). CEERS, being a lower depth survey, offers relatively low accuracy and precision when determining object redshift, and thus a high-redshift sample derived from it will have a lower purity. However, with the large and less costly nature of the survey, astrometric use cases could be employed, with \texttt{A22} exhibiting the lowest bias of all measurements. The sets \texttt{T22} and \texttt{S23} are not included in this analysis because they require pre-selection of high-redshift candidates.}
\end{table*}

Overall, most of the template sets designed for high-redshift galaxies are broadly successful at using photometry to predict spectroscopic redshifts. Here, they are compared using three datasets with varying survey designs: UNCOVER, with the deepest observations (due to its lensed field) but with fewer bands; JADES, which is not as deep as UNCOVER but has wider photometric coverage; and CEERS, a wide-band but relatively shallow survey, making it the best probe for galaxies near detection thresholds. For all three, the redshift subsample only consists of galaxies with spectroscopic $z > 5$ since the \texttt{S23} template sets preferentially do not include templates that would be valid for older (lower redshift) galaxies (subsequently reducing the number of fit parameters at higher redshift).

Previous evaluations have typically focused on the catastrophic error rate, $\eta\equiv \frac{N\left(\left|\Delta z\right|>0.15\right)}{N}$ \citep{Hildebrandt_2010}. A template set that minimizes the catastrophic error rate is then considered the most useful. However, the template sets considered here have two potential failure modes. A catastrophic error in redshift is what might be termed a silent error since there is no indication given that the photometric redshift is wrong. In addition, several template sets considered here have noisy error modes, in which they are unable to return any photometric redshift because there is no good fit. A noisy error is certainly preferable to a silent error since additional information allows for the selection of a high-purity sample. However, this also leads to a sample with lower completeness, and if the noisy errors are predominantly due to a specific SED shape (which arguably likely is true), quite possibly a biased sample as well. Thus, here two metrics are used, $\eta_f$, the catastrophic error fraction in the sample that returns a photometric redshift, and $\eta_c$, which includes both silent and noisy errors and is given as a fraction of the complete sample. This distinction is greatest in the two control samples, \texttt{Ev3} and \texttt{TwSPS}, as they were not designed for high-redshift fitting and thus are unable to return fits for 1-10\% of the high-redshift objects (Tables \ref{tab_zs_stats_JADES_high}, \ref{tab_zs_stats_UNCOVER_high}, \ref{tab_zs_stats_CEERS_high}).

\begin{figure*}[p]
    \centering
    \includegraphics[width=0.83\textwidth]{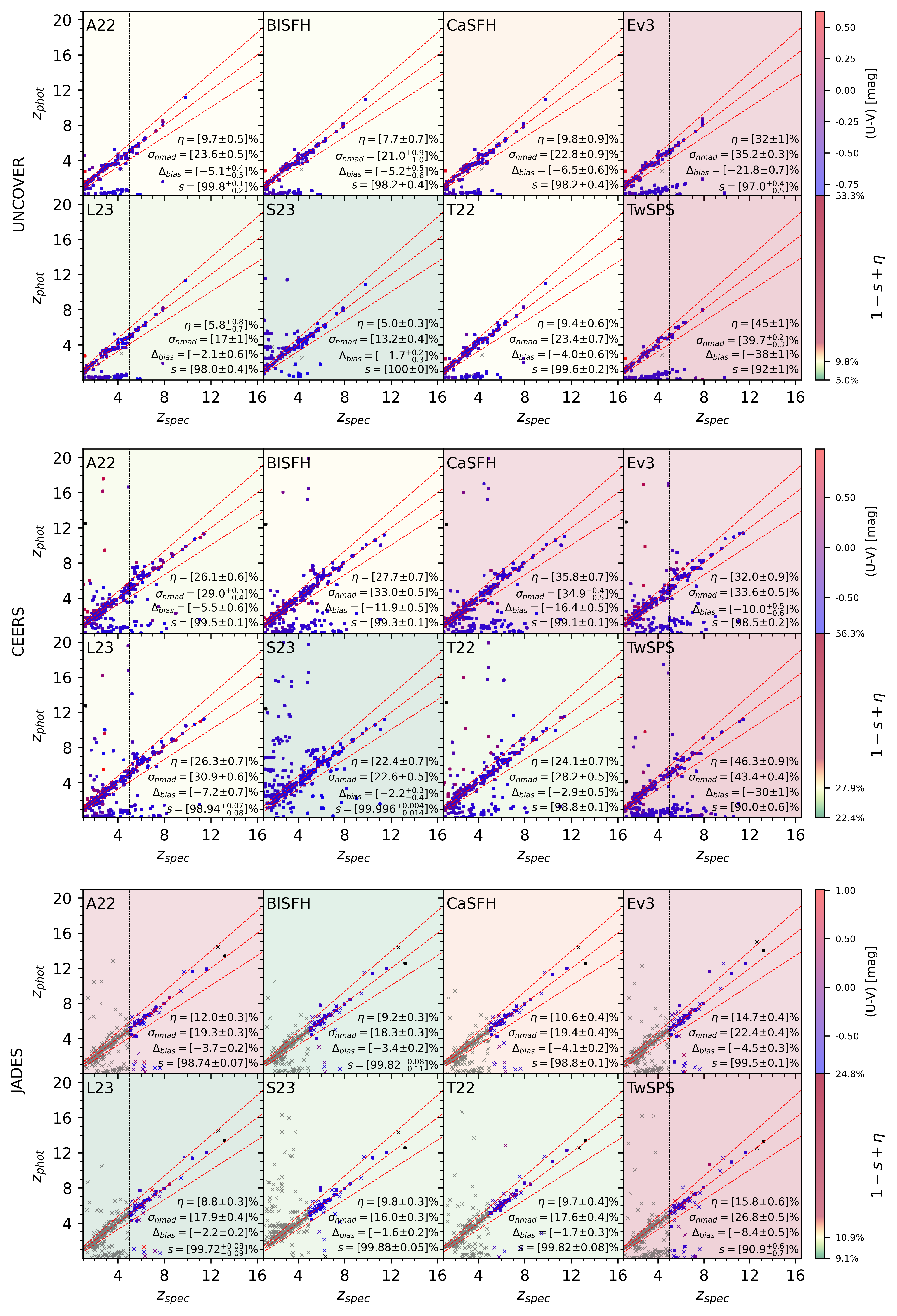}
    \caption{\label{fig:mosaic_all} JADES, CEERS, and UNCOVER template set redshift performance mosaics. The template sets are set against each other in order to compare their accuracy, completeness, and bias in the high-redshift domain ($z_{spec}>5$) through $\eta$, $\sigma_{nmad}$, $\Delta_{bias}$, and $e$. Color coding is based on the relative $1-s+\eta_f$ value, with the color scale centered on the median. Points are colored from blue to red after its color index $(U-V)$ in the rest frame. The point markers designate filter coverage, with  "$\times$" marking objects lacking four of the five NIRCam filters otherwise in the catalog, "$\circ$" marking objects with next to full NIRCam filter coverage, and "$\bullet$" marking full NIRCam coverage, along with additional MIRI filters.}
\end{figure*}

For UNCOVER (Fig. \ref{fig:mosaic_all}), the \texttt{S23} ($\eta_f = 5.0\%$), and \texttt{L23} (5.5\%) template sets produce the highest-purity redshifts (Table \ref{tab_zs_stats_UNCOVER_high}). Furthermore, \texttt{S23} is shown to have the smallest $\sigma$ and $\Delta$ as compared to other template sets. With a complete sample, \texttt{S23} is thus best for characterizing redshifts in a UNCOVER-like sample. Most of the other template sets with small silent error fractions are unable to return photometric redshifts for 1-3\% of the full sample.

For most template sets, the difficulty in constraining photometric redshifts comes from insufficient wavelength coverage rather than from insufficient depth. As a result, with the additional filters available in JADES (Fig. \ref{fig:mosaic_all}), most template sets now return fits for 99\% or more of the available sample (Tables \ref{tab_zs_stats_JADES_high}, \ref{tab_zs_stats_JADES_all}). Still, the effectively reduced depth means that most template set performs worse than for UNCOVER. \texttt{L23} ($\eta_f = 8.8\%$), \texttt{BlSFH} (9.2\%), and \texttt{T22} (9.7\%) have significantly lower catastrophic error rates and are thus preferred (Table \ref{tab_zs_stats_JADES_high}). However, \texttt{S23} (with $\eta_f=9.8\%$) has the smallest RMS error while maintaining high purity. On the entire sample, \texttt{L23} again has the smallest $\eta_c = 9.1$\%, although \texttt{S23} ($\eta_c = 9.9$\%) still has the smallest RMS error with a less significant difference in $\eta$, and therefore may be preferred. 

Finally, CEERS is toughest to fit, as every template set has a catastrophic error rate over 22\% (Tables \ref{tab_zs_stats_CEERS_high}, \ref{tab_zs_stats_CEERS_all}). Despite the decreased depth, fit return rates are relatively high ${\sim}99\%$ (Table \ref{tab_zs_stats_CEERS_high}). However, all RMS redshift errors are ${\geq}22\%$. \texttt{S23} produces the best performance by every metric, returning a fit for every object with $\eta_f \approx \eta_c = 22.4$\%. \texttt{T22}, with  $\eta_f = 24.1\%$ and $\eta_c = 25.3\%$, is the next best choice for a high-purity and completeness sample. Still, no template performs satisfactory for the full CEERS sample. The results for all three datasets are summarized in Table \ref{tab:templatesummary_z}.

Given the relatively poor fits, it is worth examining differentiating factors in the CEERS survey. Examining the band coverage (Fig. \ref{fig:filtercoverage}), the inclusion of the MIRI bands stands out. This suggests that the inclusion of MIRI-bands has little to no effect on redshift constraint at this depth. When comparing to JADES, CEERS have many fewer measured bands in the NIR spectrum. Lastly, CEERS has for the most part marginally or significantly lower depths than both UNCOVER and JADES. In summary, significant factors that improve JADES and UNCOVER's redshift estimation as compared to CEERS, include their higher depth in optical and NIR ranges, and higher high-depth filter-count.

\begin{table}[hbt!]
    \renewcommand{\arraystretch}{1.3}
    \caption{Most successful templates for photometric redshift in different scenarios.}
    \centering
    \begin{tabular}{c|cccc}
        Catalog & Best for Purity & Best for Completeness \\
        \hline
        UNCOVER & \texttt{S23} or \texttt{L23} & \texttt{S23} \\
        JADES & \texttt{L23} & \texttt{S23} or \texttt{L23}  \\
        CEERS & \texttt{S23} & \texttt{S23} \\
    \end{tabular}
\label{tab:templatesummary_z}    
\end{table}

\subsection{Asymmetry in catastrophic redshift errors}
Catastrophic redshift errors comes in two types: lower redshift objects scattering to higher redshift and higher redshift objects scattering to lower redshift. Since a common cause of catastrophic errors is an inability to distinguish between Lyman and Balmer breaks \citep{Hovis-Afflerbach_2021, Pirzkal_2013}, or low-redshift strong line emitters mimicking high-redshift breaks in photometric data \citep{Atek_2011,ref_Penin}, these errors commonly result in objects being included in or excluded from high-redshift samples. Thus, for a purely photometric study, scattering an object up, would result in the overestimation of the luminosity and mass functions at high redshifts, and similarly an object mistakenly excluded results in the underestimation of such functions. For studies with planned follow-up spectroscopy, false high-redshift targets result in wasted observing time, whereas objects scattered to low redshifts result in incomplete and likely biased samples. Depending upon the specific study, either type of error could be more important to avoid.

A strong asymmetry exists due to relative population sizes, as there are far more galaxies at $z \sim 3{-}5$ than $z > 10$ which are at the detection limit. For example, if 5\% of each population were misclassified, this would result in a high-redshift selection that would be 95\% complete; yet, it would also be dominated by low-redshift interlopers due to their relative abundances. An extreme version of this effect occurred in very early JWST studies, in which several $z \gtrsim 15$ candidates were selected from photometry, yet none of them was found to lie at these extreme redshifts in follow-up observations \citep{Yan_2022, Castellano_2022, bunker2024jadesnirspecinitialdata}. Several fitting techniques place a Bayesian prior on the redshift distribution in an attempt to mitigate this effect (including a luminosity prior in EAZY). However, an incorrect prior will produce a biased sample, and an accurate prior is typically only possible at redshifts where complete statistical samples already exist.

\subsubsection{Redshift overestimation}
The most common class of high-redshift interlopers are seen in CEERS at $z_{spec}\sim 4{-}5$, where a Balmer break in dusty star-forming galaxies can be mistaken for a Lyman break, leading to a significantly overestimated $z_{phot}$. Several of these easily mistakable objects have previously been reported \citep{Zavala_2023}, where deep millimeter-wave observations have been used to adjust redshifts of CEERS observations. However, the lack of similar interlopers in the JADES and UNCOVER suggests that the additional observational depths and filters in these surveys is the key to avoiding this type of catastrophic error. Specifically, higher depth in the $\SI{0.5}{\mu m}$ to $\SI{20}{\mu m}$ bands (${\sim} \SI{0.09}{\mu m}$ to $\SI{0.36}{\mu m}$ in rest-frame) is a noteworthy survey differentiator (especially in JADES). Additionally, the unique addition of Hubble F435W and NIRCam F090W among UNCOVER and JADES (see Fig. \ref{fig:filtercoverage}) could have significantly assisted in breaking the Lyman and Balmer break degeneracy.

Additional catastrophic redshift overestimates are common at $z_{spec}<5$ for all catalogs (Fig. \ref{fig:mosaic_all}, Tables \ref{tab_zs_stats_JADES_all}, \ref{tab_zs_stats_UNCOVER_all}, \ref{tab_zs_stats_CEERS_all}). Although there is no singular nor common cause that would produce $z_{phot} \gg z_{spec}$. However, for these objects as well, it is likely that additional observing time or different observing strategies would mitigate such catastrophic redshift overestimations the need for spectroscopy. 

For JADES, these catastrophic overestimates only occur in objects without full filter coverage. This illustrates the importance of broad SED coverage to avoid confusion between the emission lines and spectral breaks. CEERS, with similar filter coverage at lower depths, contains overestimates even for objects with full coverage. Thus, it is a combination of depth and coverage that ensures low catastrophic redshift overestimation rates. Specifically, one can conclude that the JADES\&JEMS $>20$ filters with depth ${\gtrsim}29$ should be sufficient to produce a high-purity photometric sample of high-redshift galaxies.

\begin{table*}[t]
\renewcommand{\arraystretch}{1.3}
\caption{\centering Fit residuals for the SED sample on both photometric and spectroscopic data.}
\label{tab:sed_chi2}
\centering
\small 
\begin{tabular}{ccccccccccccccccc}
\hline\hline
NIRCam ID & \multicolumn{2}{c}{\texttt{A22}} & \multicolumn{2}{c}{\texttt{BlSFH}} & \multicolumn{2}{c}{\texttt{CaSFH}} & \multicolumn{2}{c}{\texttt{Ev3}} & \multicolumn{2}{c}{\texttt{L23}} & \multicolumn{2}{c}{\texttt{S23}} & \multicolumn{2}{c}{\texttt{T22}} & \multicolumn{2}{c}{\texttt{TwSPS}} \\ 
\cline{2-17}
 & $\chi^{2}_{s}$ & $\chi^{2}_{p}$ & $\chi^{2}_{s}$ & $\chi^{2}_{p}$ & $\chi^{2}_{s}$ & $\chi^{2}_{p}$ & $\chi^{2}_{s}$ & $\chi^{2}_{p}$ & $\chi^{2}_{s}$ & $\chi^{2}_{p}$ & $\chi^{2}_{s}$ & $\chi^{2}_{p}$ & $\chi^{2}_{s}$ & $\chi^{2}_{p}$ & $\chi^{2}_{s}$ & $\chi^{2}_{p}$ \\ 
\hline
110996 & 5.5 & 7.8 & 4.4 & 5.4 & 4.4 & 5.4 & 2 & 2.2 & 1.1 & 3.8 & 12 & 15 & 1 & 1.3 & 7.8 & 8.6\\
132780 & 6.9 & 7.8 & 9.1 & 23 & 7 & 7.3 & 6.6 & 34 & 2.8 & 3.4 & 9.9 & 20 & 3.4 & 5.3 & 8.4 & 9.2\\
110319 & 2.9 & 2.9 & 4.5 & 11 & 1.6 & 1.9 & 3.3 & 14 & 1.5 & 1.8 & 4.4 & 9.1 & 1.8 & 1.9 & 3.3 & 3.4\\
113585 & 110 & 110 & 140 & 330 & 74 & 48 & 110 & 1100 & 59 & 50 & 100 & 210 & 62 & 54 & 100 & 100\\
106292 & 9.5 & 11 & 12 & 22 & 12 & 18 & 13 & 29 & 4.2 & 4.8 & 17 & 28 & 4.2 & 3.9 & 13 & 13\\
106885 & 6.5 & 8.1 & 7.5 & 23 & 5.8 & 19 & 4.4 & 18 & 2.7 & 5.4 & 10 & 24 & 2.5 & 3.6 & 8.1 & 9\\
\hline
\end{tabular}
\vspace{0.3cm}
\tablefoot{Overview of fitting performance when comparing the reconstructed SED to NIRSpec observations for the SED sample. The results solely using photometric observations ($\chi^2_{\nu,phot}=\chi^2_{p}$) as well as fits to the full combination of photometry and spectroscopy ($\chi^2_{\nu,spec}=\chi^2_{s}$) are compared to indicate how well the observed SEDs are represented in each template set. We note that the errors used to calculate ($\chi^2$) for each case come from the observed SEDs, and thus rely on the accuracy of the error estimate in these datasets. Although a few objects such as 110996 and 110319 are relatively well fit by some templates, others (e.g., 113585) seem entirely unrepresented in current template sets. It should be noted that $\chi^2_{spec}$ is occasionally greater than $\chi^2_{phot}$, which is ideally not to be expected. This is most likely associated with potential minor nonlinearities in the EAZY fitting procedure that can lead to slightly approximated solutions (G. Brammer, private communication).}
\end{table*}

UNCOVER has a distinct lack of similar catastrophic redshift overestimations for all $z_{spec}>1$ despite offering a lower number of filters as compared to JADES. With the main other relative difference between the two surveys being effective depth, it suggests that the additional depth also provides sufficient constraint to hinder catastrophic redshift overestimation. The additional effective depth of a lensed field is thus an effective solution to benefit purity in high-redshift selection.

\subsubsection{Catastrophic redshift underestimation}
The same confusion between Balmer and Lyman breaks also causes high-redshift galaxies to be fit at $z_{phot} \ll z_{spec}$. The problem is exacerbated because the highest-redshift objects in any survey, regardless of depth, always lie near the detection threshold, producing low signal-to-noise ratios. Further, for the highest-redshift objects, fewer bands are available with non-zero detections due to the longer observed wavelength of the Lyman break. 

Technically, this would be avoided by using template sets that do not contain galaxies with older stellar populations or high extinction since any discernible break would only be fit as a Lyman break by such templates. Thus, by construction, there are fewer catastrophic redshift underestimates produced especially by the \texttt{S23} template set. However, the same lack of templates with Balmer breaks would result in an overwhelming population of catastrophic redshift overestimates if used without a prior high or redshift assessment. Using \texttt{S23} as an example, it is clear to see that these young-population templates only are valid for use after a prior demarcation of whether objects are high redshift or not.

Overall, the best procedure for determining photometric redshifts of high-redshift galaxies should therefore be to combine two template sets. First, \texttt{L23} should be run to produce an initial selection of high-redshift candidates. Of the template sets that include older stellar populations and lower redshift models, it has the fewest catastrophic redshift underestimates. Therefore, it produces the most complete sample of high-redshift candidates. Subsequently, \texttt{S23} should be run on that restricted sample.

\begin{figure}
    \centering
    \includegraphics[width=1\linewidth]{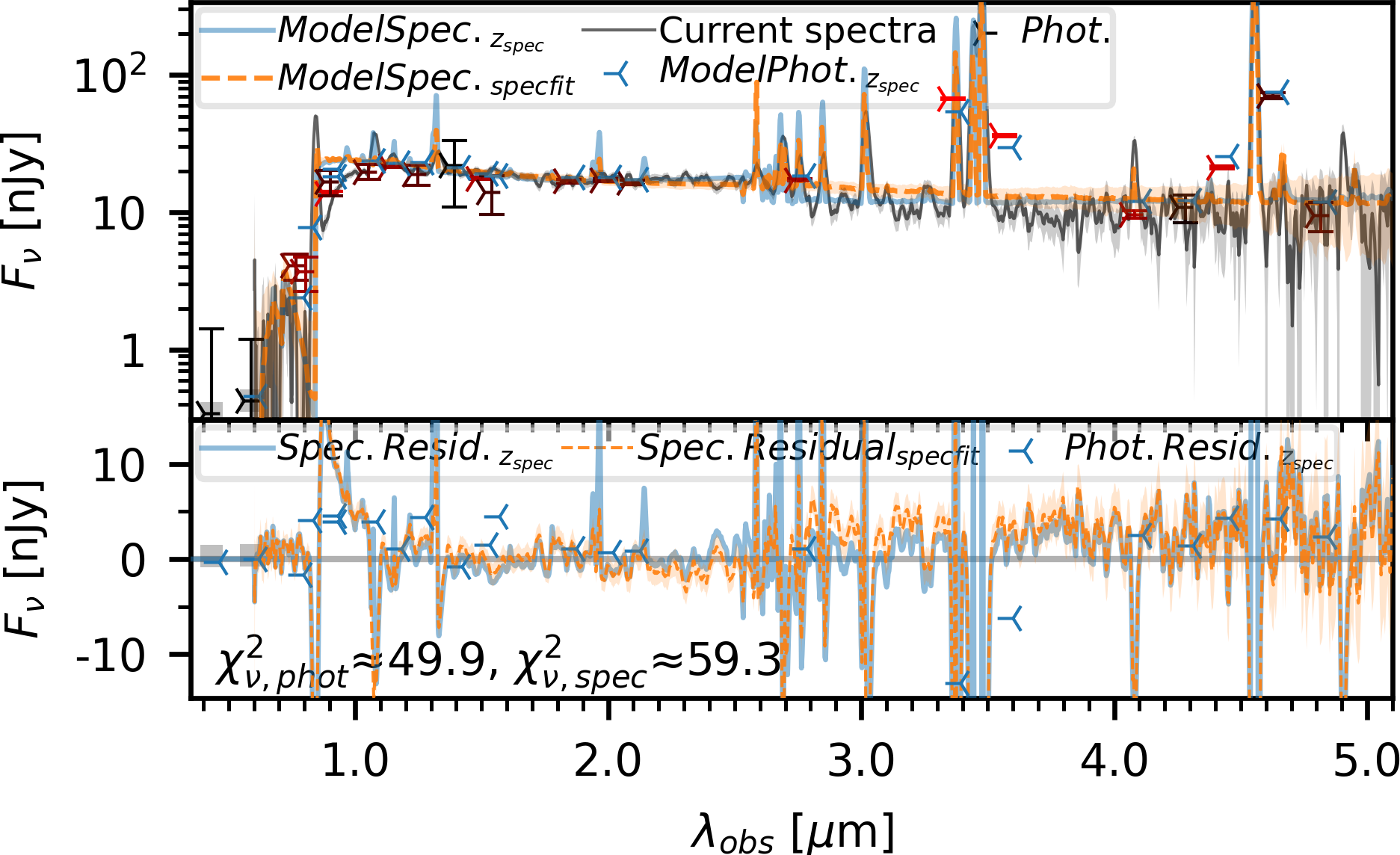}
    \caption{\label{fig:L23_113585}Best-fit \texttt{L23} reconstruction for NIRCam-ID=113585, with $z_{spec}=5.9$ and $z_{phot}=5.9$. In this fit, $z$ is constrained to $z_{spec}$. The template is fit both to photometry (in blue) and a concatenated photometric and spectroscopic dataset (orange). JADES 113585 appears to be poorly represented in template sets optimized for high-redshift galaxies, likely due to emission lines and Lyman break physics.
   }
\end{figure}

\subsection{SED analysis}
\label{sec:sedanalysis}
The templates are also evaluated on their ability to reproduce the full observed SED using the six objects from the SED sample (Fig. \ref{fig:specobjs}). Although this is a small sample, the test itself is critical because in order for the properties derived using models to be reliable, these models must satisfactorily reproduce the full SED. If uncertainties are properly calibrated and independent, this would result in a $\chi^2_{\nu} \leq 1$, with respect to either the spectra or the photometry.

For photometry, this is indeed what results from template fitting. For five of the six objects in the SED sample, the best-fit reconstructed SEDs have $\chi^2_{\nu} \leq 1$ compared against the photometry. This demonstrates that the photometric information has been entirely exhausted. The test presented here determines whether these reconstructions are good approximations to the true SED or whether the additional information from spectroscopy is necessary to produce a meaningful fit.

\begin{figure}[ht]
    \centering
    \includegraphics[width=1\linewidth]{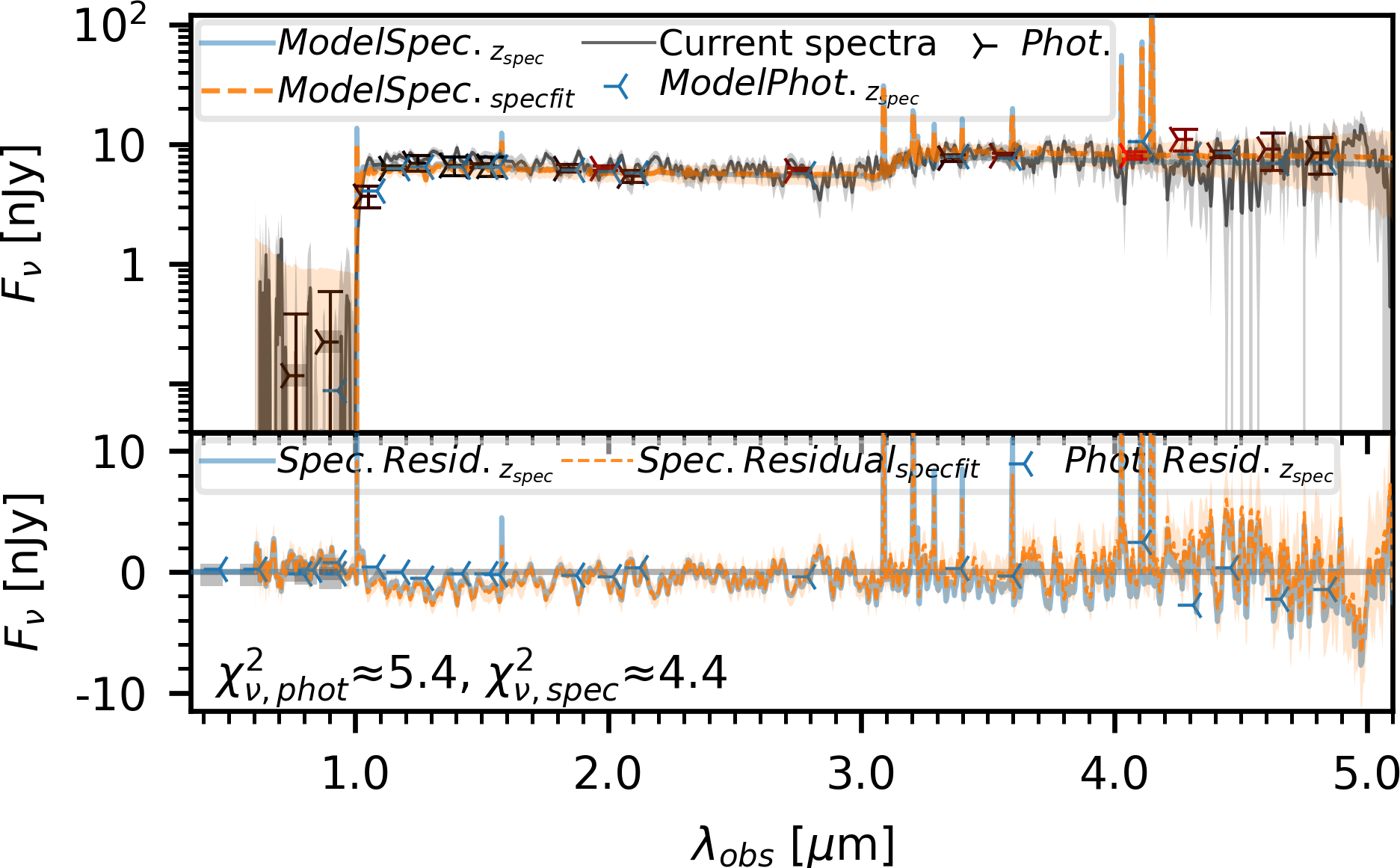}
    \caption{\label{fig:BlSFH_110996}Best-fit \texttt{BlSFH} reconstruction for NIRCam-ID=110996, with $z_{spec}=7.3$ and $z_{phot}=7.7$. In this fit, $z$ is constrained to $z_{spec}$. The template is fit to both to photometry (in blue) and a concatenated photometry and spectroscopy dataset (orange). The template produces significantly stronger emission lines than in the observed SED of JADES 110996 since there are no \texttt{BlSFH} templates with minimal emission lines.}
\end{figure}

\begin{figure}
    \centering
    \includegraphics[width=1\linewidth]{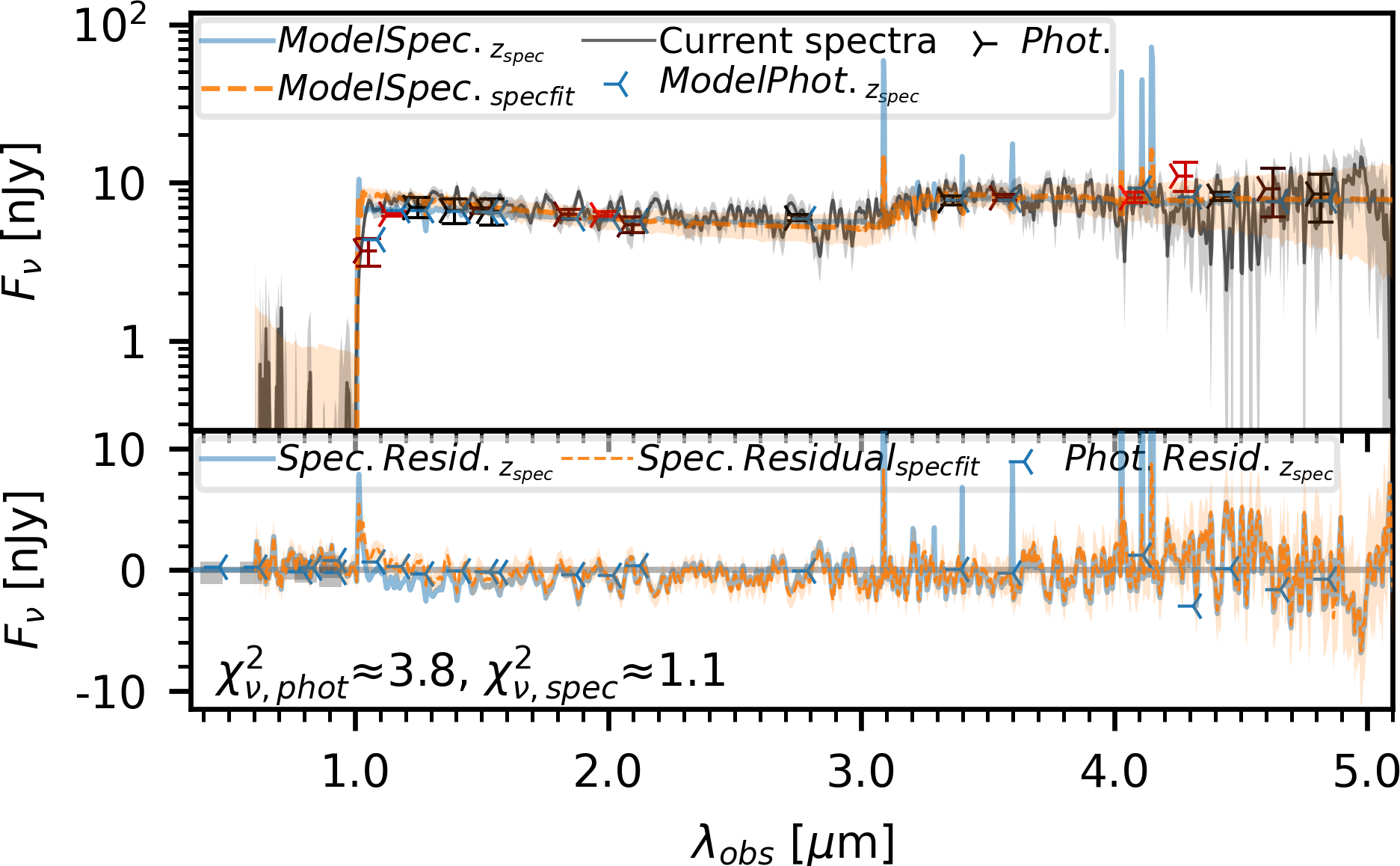}
    \caption{\label{fig:L23_110996}Best-fit template with \texttt{L23} on object NIRCam-ID=110996, with $z_{spec}=7.3$ and $z_{phot}=7.7$. In this fit, $z$ is constrained to $z_{spec}$. The template is fit to both photometry (in blue) and a concatenated photometry and spectroscopy dataset (orange). ID denotes the JADES catalog photometric ID, and $\chi^2_\nu$ for "spec" and "phot" refers to which data input is used.
    Here again, the largest discrepancies are due to overestimated emission lines, and to a minor degree infrared features.}
\end{figure}

The overall performance is summarized in Table \ref{tab:sed_chi2}. Although several templates have comparable performance, \texttt{L23} and \texttt{T22} produces the closest match to the observed spectroscopy. The \texttt{S23} template, which was the most successful at producing complete samples of high-redshift galaxies from photometry, is unable to successfully describe any of the six full spectroscopic SEDs. \texttt{BlSFH} has similar behavior, if slightly less extreme. 

\begin{figure}
    \centering
    \includegraphics[width=1\linewidth]{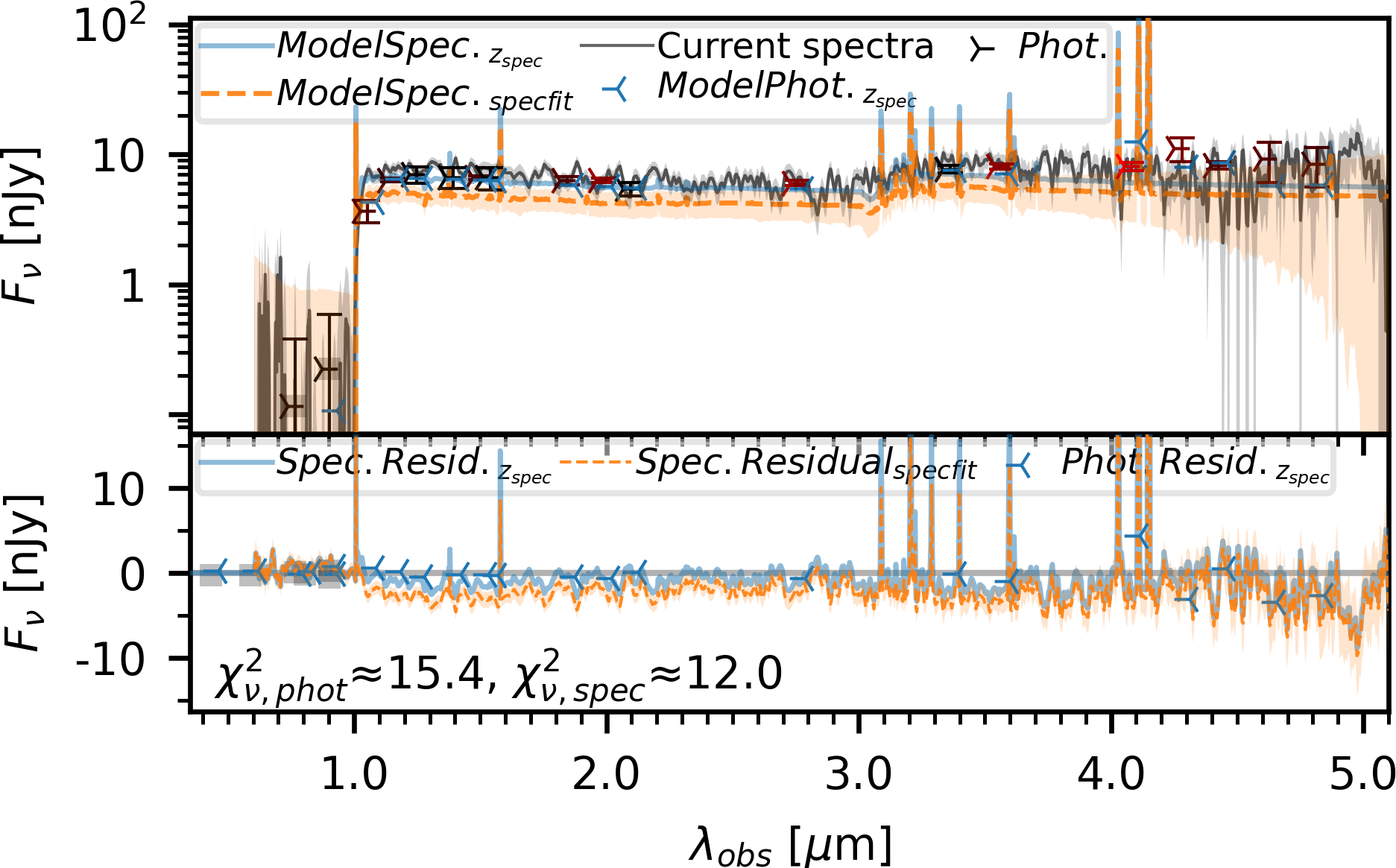}
    \caption{\label{fig:S23_110996}Best-fit template with \texttt{S23} on object NIRCam-ID=110996, with $z_{spec}=7.3$ and $z_{phot}=7.8$. In this fit, $z$ is constrained to $z_{spec}$. The template is fit to both photometry (in blue) and a concatenated photometry and spectroscopy dataset (orange). ID denotes the JADES catalog photometric ID, and $\chi^2_\nu$ for "spec" and "phot" refers to which data input is used.
    Here again, the largest discrepancies are due to overestimated emission lines, but in this fit, the minimization of $\chi^2_\nu$ caused the overall features to be under-scaled.}
\end{figure}

\texttt{S23} is unique in its treatment of the high-redshift IMF but does not produce a consistent set of corresponding changes to nebular emission. Nebular emission has been observed to be significantly stronger than anticipated at high redshift \citep{Carnall_McLure_Dunlop_McLeod_Wild_Cullen_Magee_Begley_Cimatti_Donnan_et_al._2023}, so a failure to model the nebular emission properly leads to an inability to match the full SED. However, the salient features used to produce photometric redshifts instead rely on the stellar population and its age, which is improved in the astrophysical model used by \texttt{S23}. An ideal template set would therefore combine the improved models for both the stellar population and nebular emission in a consistent manner.

For most models, the reduced $\chi^2_{\nu}$ between observed spectroscopy and the SED reconstructed from photometry alone is similar to the best-fit $\chi^2_{\nu}$ given the full spectrum. Thus, photometry alone is generally sufficient to produce the best-fit reconstruction with a template. Further, EAZY is able to successfully find the right solution within a set of templates, even if that template set does not always produce a good description of the true SED.

If this remains true as models improve, it would produce a surprising and optimistic conclusion. For more local galaxies, it has been well established that multi-wavelength observations are necessary to robustly determine the best reconstructed SEDs \citep{10.1093/mnras/stu2195,Förster_Schreiber_2004, 10.1093/mnras/stv1727}. As a result, broad wavelength coverage is required to constrain stellar masses, star formation rates, and other physical parameters. However, at high-redshift there is a more limited set of bands with non-zero detections. If these fewer bands are sufficient to robustly determine astrophysical parameters, then purely photometric surveys with JWST will be sufficient to produce mass functions and related studies.

However, at present the most successful models still yield fits with a median $\chi^2_{\nu} \sim 2-5$, suggesting that even the best current models still are unable to accurately describe the full SED. For object 113585 (NIRCam ID, MAST), the mismatch is particularly strong, with no model producing $\chi^2_{\nu}$ $\lesssim 50$. This appears to be part of a class of extremely blue object that have features not represented in any of the template sets. The \texttt{L23} reconstruction is a good match for the rest-frame UV emission (Fig. \ref{fig:L23_113585}). However, this is with the strong exceptions of the Lyman break, and the strong \ce{Ly$\alpha$}$\lambda121.6$ ($=\SI[separate-uncertainty = true]{1035(60)e-19}{\Jys}$) and \ce{[C IV]}$\lambda154.9$ ($=\SI[separate-uncertainty = true]{386(30)e-19}{\Jys}$) lines \citep{bunker2024jadesnirspecinitialdata}. In the redder end, emission lines are seen to be stronger in the residuals without consistent over or under estimations between the two fits. The continuum emission is overestimated in the red end, with the effect stronger among the other templates (see Appendix \ref{sappendix:113585}), making the object bluer than expected by templates.

An example of a strong difference between template sets is 110996 \citep{trussler2024likecandlewindembers}, which is fit well by \texttt{T22} and \texttt{L23} but poorly by several other template sets (As seen in Figures \ref{fig:L23_110996},\ref{fig:BlSFH_110996},\ref{fig:S23_110996}). 110996 has very weak emission lines, a possibility which is only part of the \texttt{T22} template set. As a result, this object is best fit by \texttt{T22}. All other templates (including to a small degree \texttt{L23}), were forced to fit strong emission lines, leading to incorrect stellar populations and ages. 

In summary, a majority of high-redshift galaxies are currently poorly described by every template set evaluated. The size of the discrepancy between observation and model varies, both between template sets and depending upon the shape of the SED. The best template set, \texttt{L23} and \texttt{T22}, only have $\chi^2_{\nu} < 2$ for two of the six objects. Furthermore, the templates with the best performance in determining photometric redshifts did not fit any of the six SEDs successfully. Significant improvements will be needed before template fitting, can be used to derive reliable astrophysical properties. However, for the better fit cases, $\chi^2_{\nu,spec}\sim\chi^2_{\nu,phot}\sim1.5$, implying that improved models would enable extracting successful SEDs from purely photometric surveys even at high redshift.

\section{Discussion}
\label{sec:discussion}
In summary, if template sets are selected correctly and used on deep datasets, they can fit adequate models for typical high-redshift galaxies. Within the redshift sample, most templates can be trusted to only rarely produce catastrophic redshift errors since the cause of bad fits can often be attributed to poor filter coverage, depth, or both (see Figure \ref{fig:mosaic_all}). Of the templates considered, the \texttt{S23} template has roughly the best performance when constrained to the high-redshift set ($\eta\sim10$); however, it must be paired with a prior selection screen (Tables \ref{tab_zs_stats_JADES_high}, \ref{tab_zs_stats_UNCOVER_high}, \ref{tab_zs_stats_CEERS_high}). The \texttt{L23} template is generally the set that performs best in creating such a prior. It performs with a lower, but not great, $\eta\sim 10\%$. If a prior is not possible, \texttt{L23} also has the best rough estimation performance of the templates considered.

The template sets \texttt{L23} and \texttt{T22} best represent the full SED for the six high-redshift objects tested, albeit for a small sample that likely does not represent the full range of high-redshift galaxies. Even within the limited sample, it is clear that some classes of objects are not well represented by the current templates. This includes both largely featureless objects, which indicate a class of even lower metallicity objects than has been expected in these templates, and a set of extremely blue objects. Thus, some revision in the astrophysical models producing these templates appears to be required.

High-redshift templates are primarily developed using one of several approaches to choose parameters for SPS models. These can include simulations, astrophysical arguments from first principles, or even empirical constraints. In catalogs with fewer observational constraints (depth, coverage), the most successful redshift predictions come predominantly from templates adopting simple astrophysical models to generate their templates. This includes mainly \texttt{S23}, which is produced primarily from first principles. The more detailed models arising from either more complex astrophysical constraints or numerical simulations produce less reliable photometric redshifts. Thus, one might be tempted to conclude that the additional complexity introduced by these models is a poor description of the true astrophysics of the earliest galaxies.

However, improved observations produce a different picture. In the UNCOVER and CEERS high-redshift samples, \texttt{S23} and \texttt{T22} outperform other templates in redshift estimation in many metrics (especially $\eta_f$) (Tables \ref{tab_zs_stats_UNCOVER_high}, \ref{tab_zs_stats_CEERS_high}). In JADES, however, with the increased depth and coverage, \texttt{S23} no longer produces the fewest catastrophic outliers (Table \ref{tab_zs_stats_JADES_high}). A closer examination of the objects with incorrect redshifts reveals that with the increased depth in the NIR bands, these simple models are not the best performing models. As measurements improve, it is the more sophisticated models that start to produce the best redshifts.

This can be seen even more strongly when comparing reconstructed SEDs with JWST spectroscopy. The more sophisticated models such as \texttt{L23} and \texttt{T22} strongly outperform \texttt{S23}. Some of this might be due to specific missing features, such as damped Ly$\alpha$, which is not present in any \texttt{S23} template. However, more generally, this indicates that the simpler models cannot reproduce the full complexity of high-redshift SEDs. This simplicity also avoids complex degeneracies and is thus well suited for photometric redshift estimation from limited information. Still, the fits are poor predictors of the full SED and thus should not be used to estimate other astrophysical properties, such as stellar masses.

The best performance in predicting the full SED comes from models that are empirically motivated to match observations. Thus, both the simpler first principles arguments and more complex simulations appear to still be incomplete descriptions of the earliest galaxies. A natural path toward improvement then comes from modeling galaxies that are poorly represented by templates generated using these techniques.

Perhaps the most intriguing objects are those such as 113585, which is poorly fitted by every template set considered here. Even empirical models fail to fit the extremely blue spectra of 113585, suggesting that it has properties that are uncommon in previous samples at only slightly lower redshifts. One might expect that rather than merely being an individual object, 113585 could be typical of larger classes of $z > 10$ galaxies. An additional feature of the object is the strong flux from otherwise minor emission lines, with the \ce{[He I]}$\lambda587.5$ line in object 113585 being the most pronounced ($=\SI[separate-uncertainty = true]{25.6(1.5)e-19}{\Jys}$; \citealt{bunker2024jadesnirspecinitialdata}) emission, which is not prominent in any of the template sets.

It is particularly surprising that the extremely blue 113585 is not fit well by \texttt{L23}, which had an explicit goal of fitting extremely blue objects when developing the template set. Although one possible explanation is missing astrophysics, another possible explanation could still lie in the data reduction. JWST catalogs produced using different reduction pipelines still exhibit significant differences in the resulting spectroscopy after more than two years since first light (see Fig. \ref{fig:discrepancy}). Currently, the standard approach still relies on calibrating spectroscopy to photometric observations rather than on an absolute calibration.\footnote{Such as the DAWN JWST archive; \url{https://dawn-cph.github.io/dja/}} Errors in data reduction could also be responsible for the mismatch between observation and theory. If so, any template set empirically tuned to match spectroscopic observations would essentially be overfit to incorrect reductions and would need to be regenerated every time an improvement is made to the reduction pipeline.

It is also possible that templates lack other astrophysics that are significant at higher redshifts. For instance, active galactic nuclei (AGN) type sources might be surprisingly frequent at early times \citep{scholtz2025jadeslargepopulationobscured}, but because they were not expected to be common, they have generally not been included in high-redshift template sets. As such, it would be worthwhile to examine the redshift and SED fit improvement from inclusions of template spectra that include an AGN or quasar component.

Finally, it is worth noting that because no template set is representative of the full SED for even a majority of high-redshift galaxies, there is no single optimal choice. Rather, the best choice depends on the use case, survey plan, and survey depth. The best basis for producing photometric redshifts from lower-depth photometry is only suitable as the second step in a two-step procedure. Here, a prior screening of whether an object can be classified as high-redshift or not is needed. For multi-wavelength surveys with many bands and high depth, a different basis becomes optimal. And yet, a different basis is best for fitting the full SED, although no basis can fit the majority of the galaxies tested. Thus, significant improvements in both modeling and data reduction will be needed before template fitting can produce robust results with well-constrained astrophysical properties for high-redshift sources. Perhaps additional observations would be beneficial to uncover the astrophysics of objects that cannot currently be fit. Until then, only redshift and luminosity appear to be reliable properties (however, only when specific observational criteria of depth and coverage are imposed), so the primary result of high-redshift surveys are luminosity functions. The SEDs and astrophysical parameters, on the other hand, require further development in the reliability of the spectral pipeline and of our understanding of high-redshift objects to achieve a complete fit of even the current sample.\begin{acknowledgements}
The authors would like to thank Charlie Lind-Thomsen, Gabriel Brammer, Gustav Lindstad, Harley Katz, Jack Turner, Nathan Adams, Martin Rey, Ryan Endsley, and Stephen Wilkins for helpful comments, and Vadim Rusakov for code contributions. 

Some of the data products presented herein were retrieved from the Dawn JWST Archive (DJA). DJA is an initiative of the Cosmic Dawn Center (DAWN), which is funded by the Danish National Research Foundation under grant DNRF140.

The Cosmic Dawn Center (DAWN) is funded by the Danish National Research Foundation under grant no. 140. C.S. was supported by research grants (VIL16599, VIL54489) from VILLUM FONDEN.
\end{acknowledgements}

\bibliographystyle{aa}
\bibliography{refs.bib}

\begin{appendix}
\onecolumn
\section{Best SED fits}
\begin{figure*}[h]
    \centering
    \subfigure{
        \includegraphics[width=0.48\textwidth]{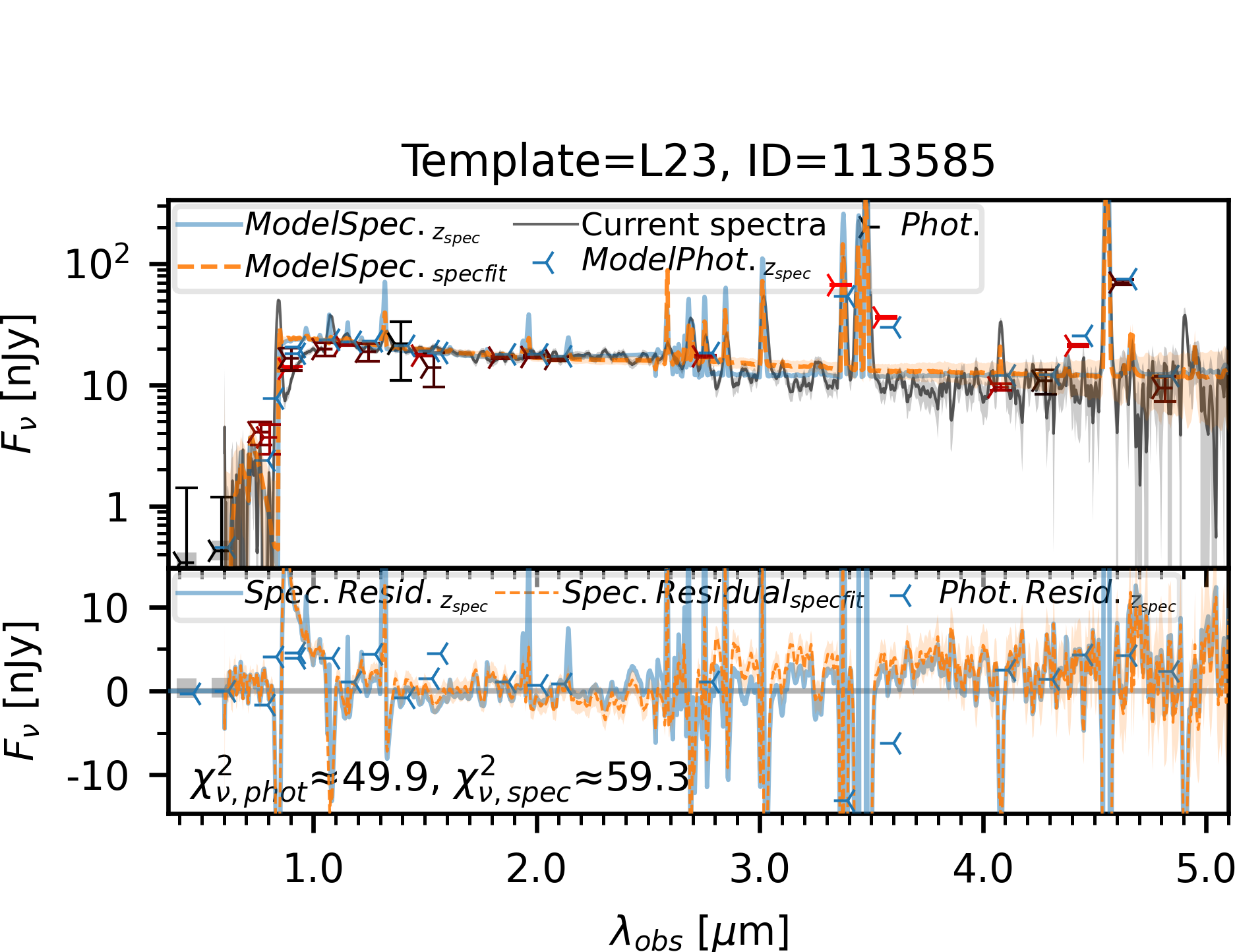}
    }
    \subfigure{
        \includegraphics[width=0.48\textwidth]{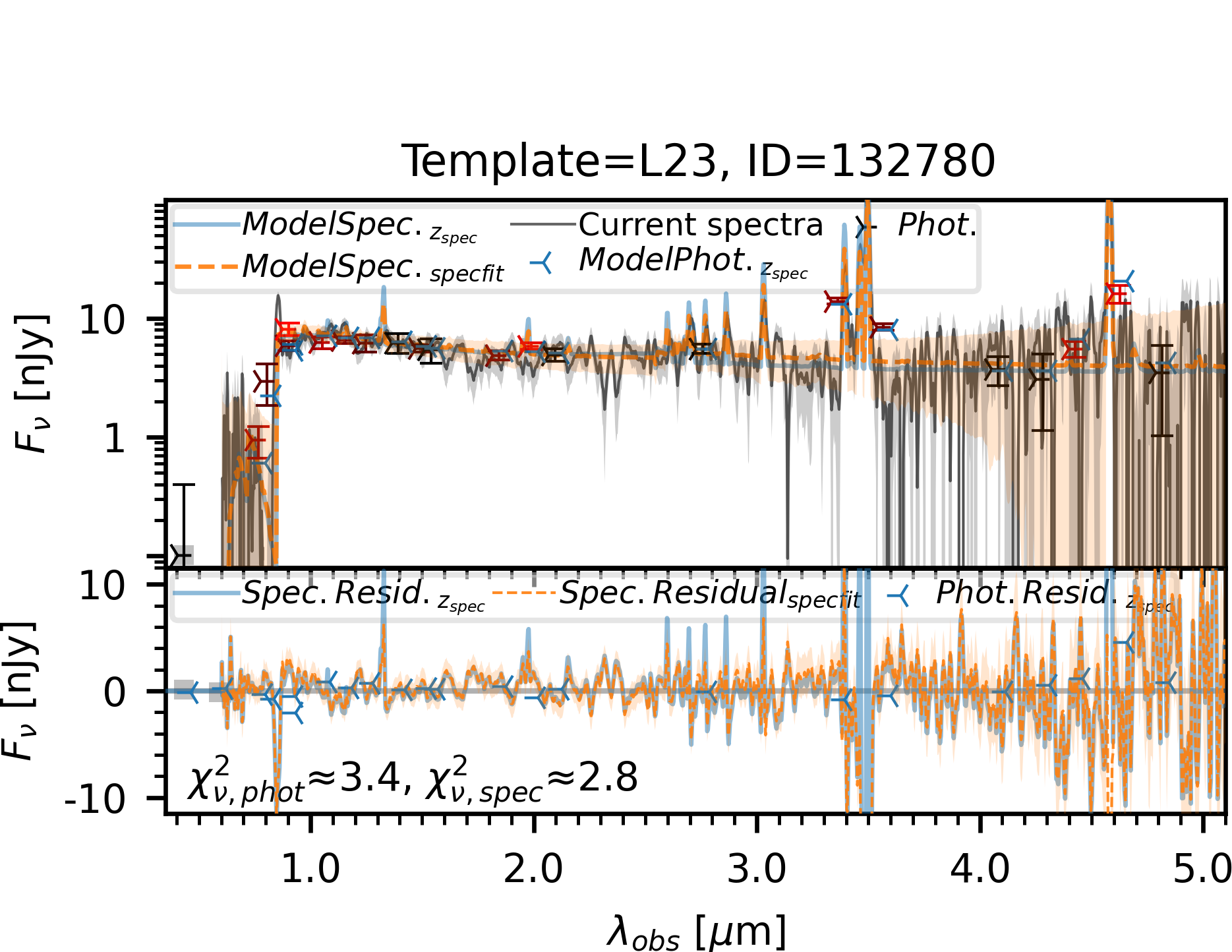}
    } 
    \subfigure{
        \includegraphics[width=0.48\textwidth]{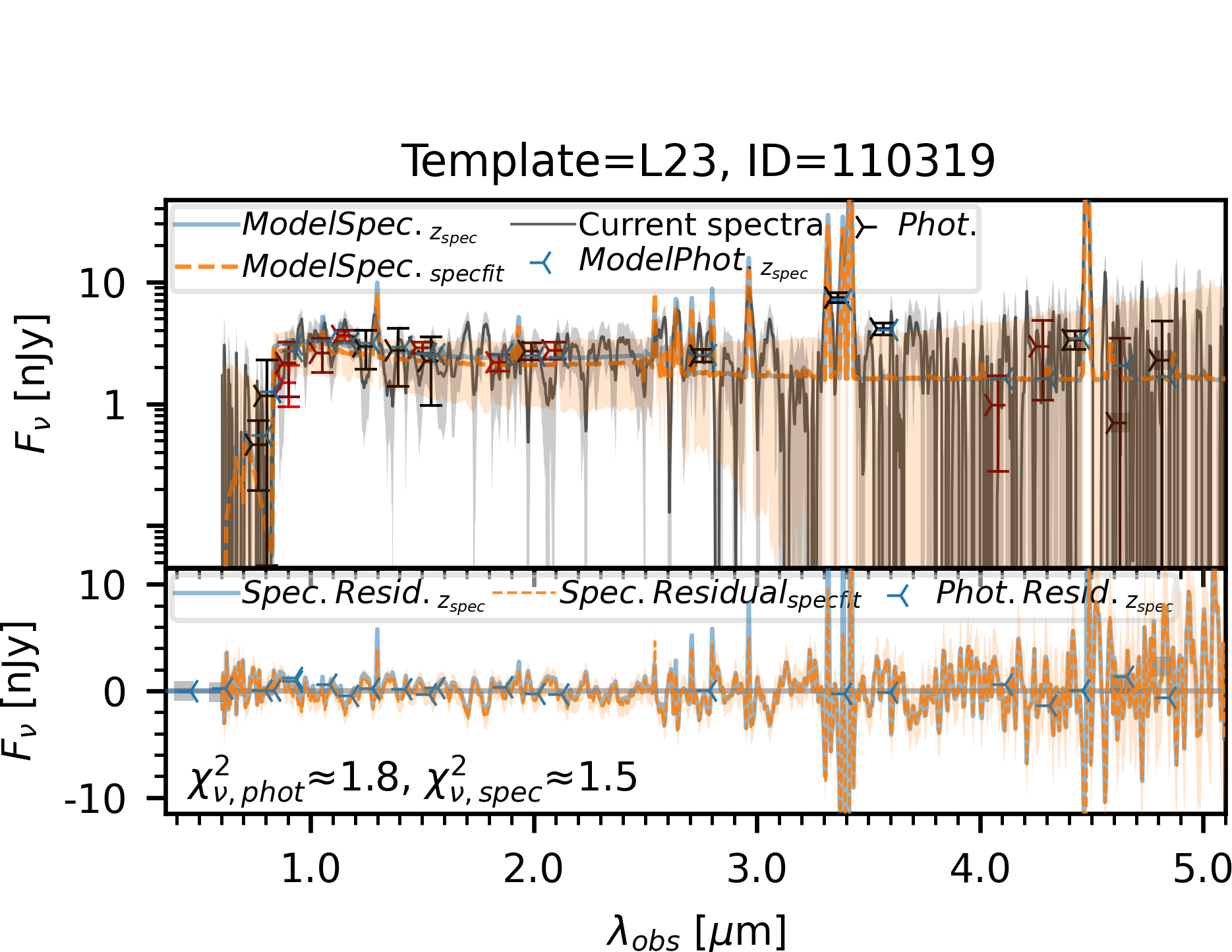}
    }
    \subfigure{
        \includegraphics[width=0.48\textwidth]{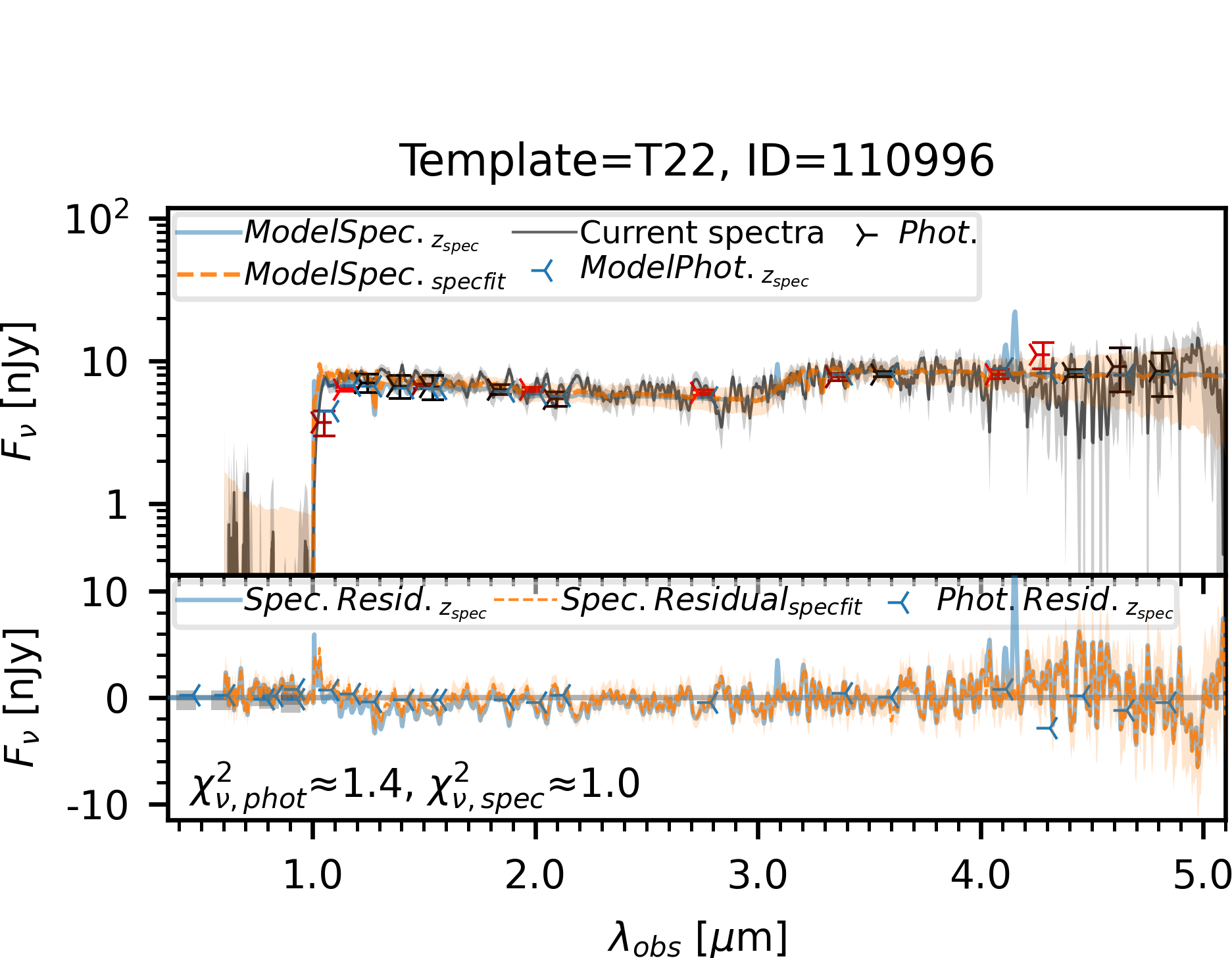}
        }
    \subfigure{
        \includegraphics[width=0.48\textwidth]{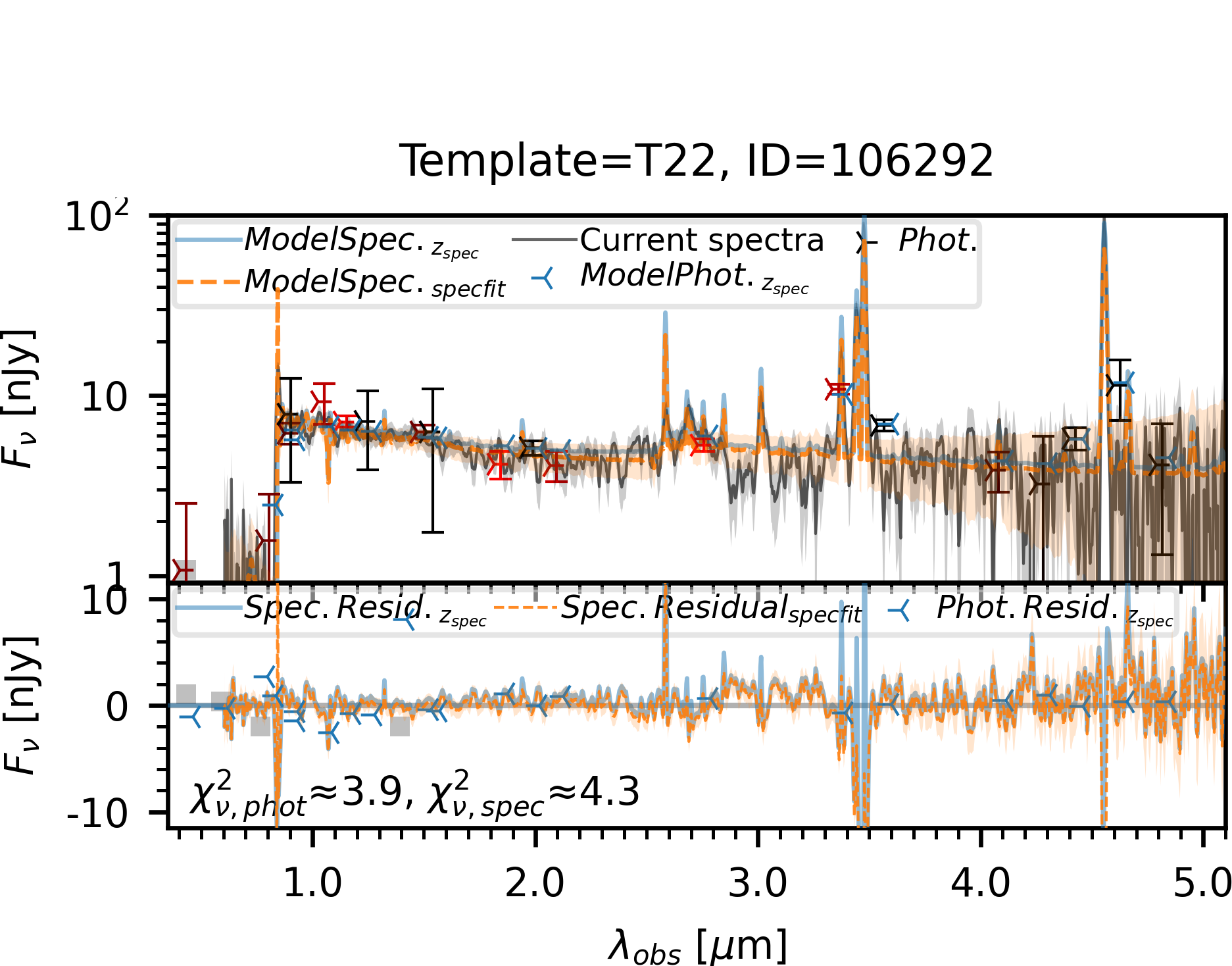}
    }
    \subfigure{
        \includegraphics[width=0.48\textwidth]{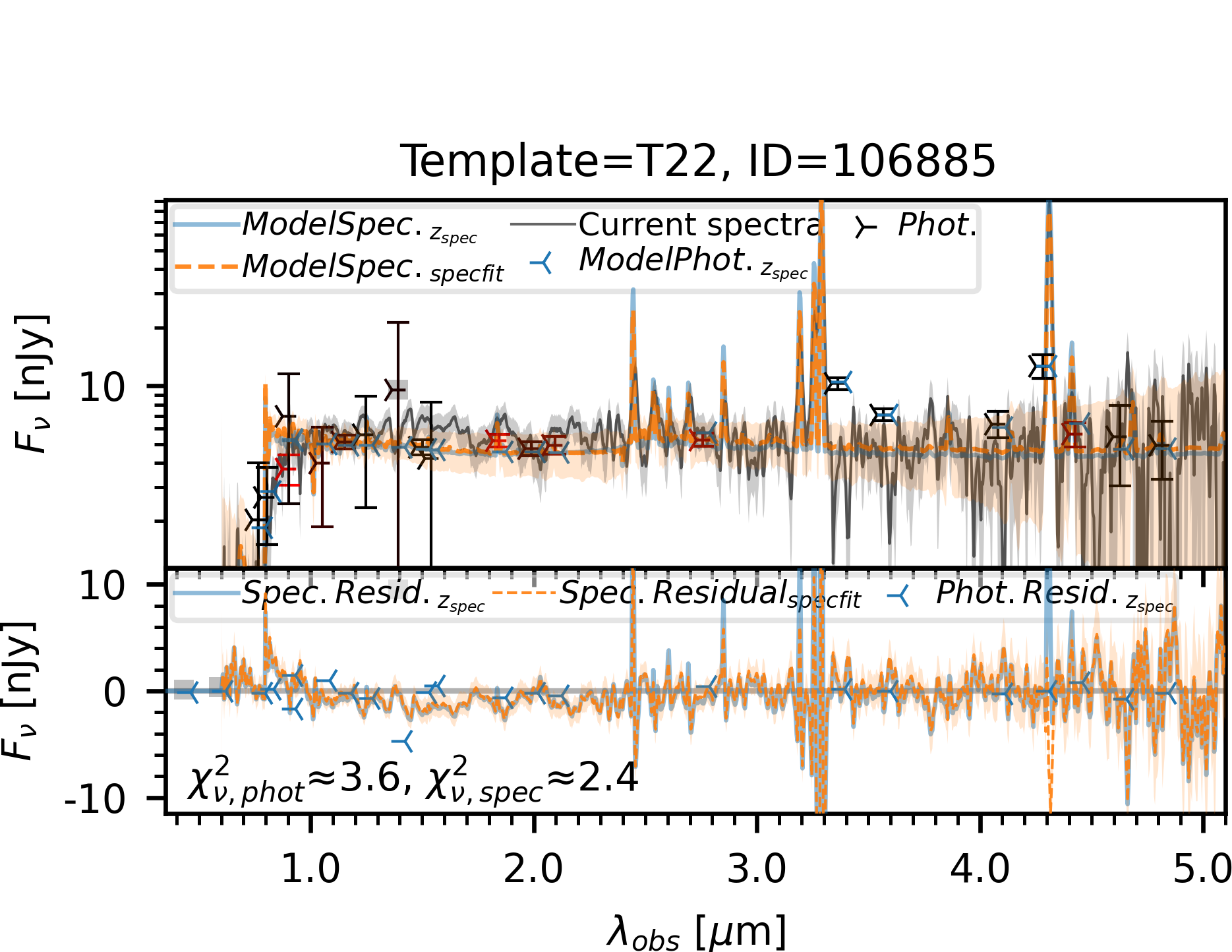}
    }
    \caption{\centering Best-fitting template fit for every object in the SED sample.}
\end{figure*}
\newpage

\section{Object 113585 fits with additional templates}
\label{sappendix:113585}
\begin{figure*}[h]
    \centering
    \subfigure{
    \includegraphics[width=0.48\textwidth]{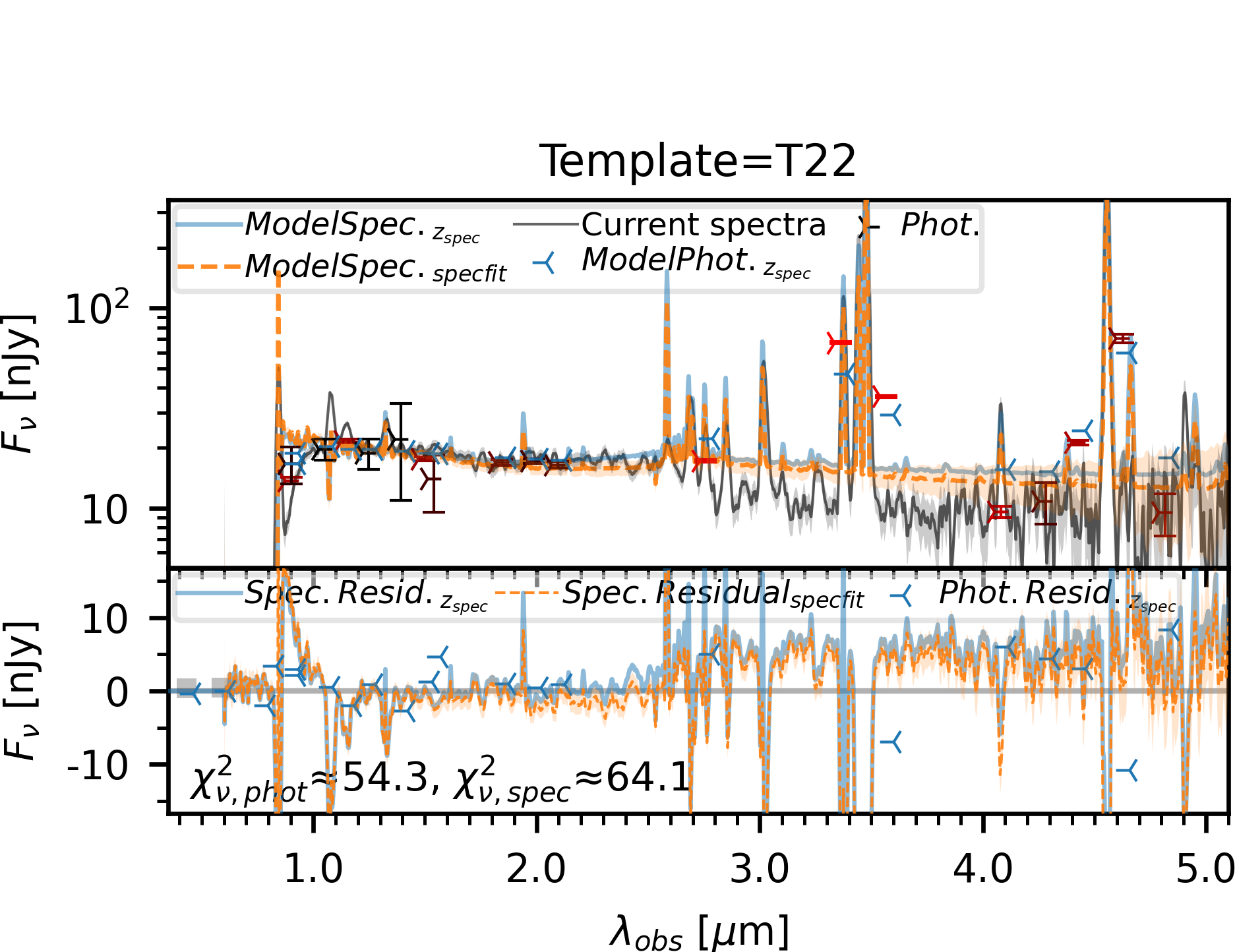}
    }
    \subfigure{
        \includegraphics[width=0.48\textwidth]{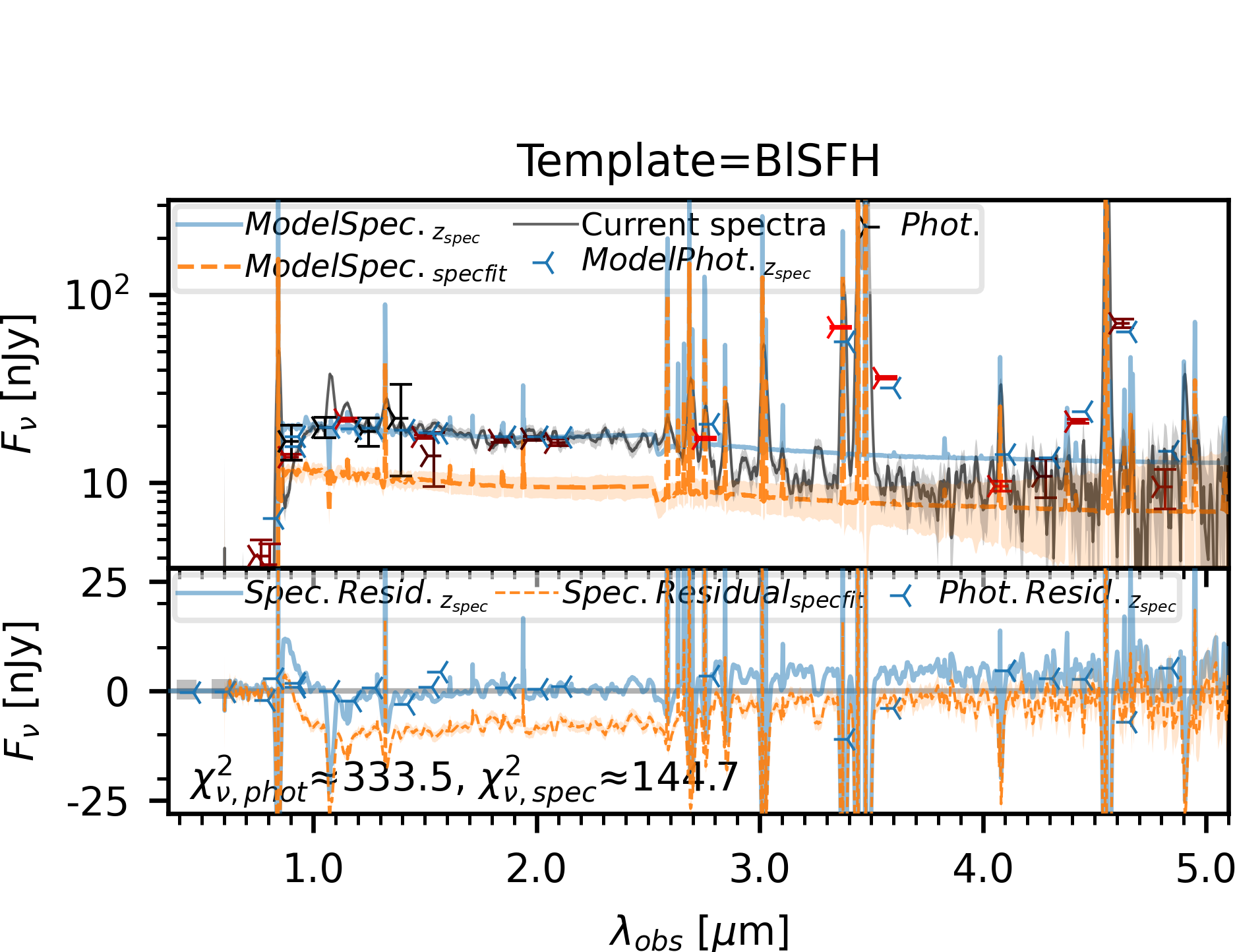}
    } 
    \subfigure{
        \includegraphics[width=0.48\textwidth]{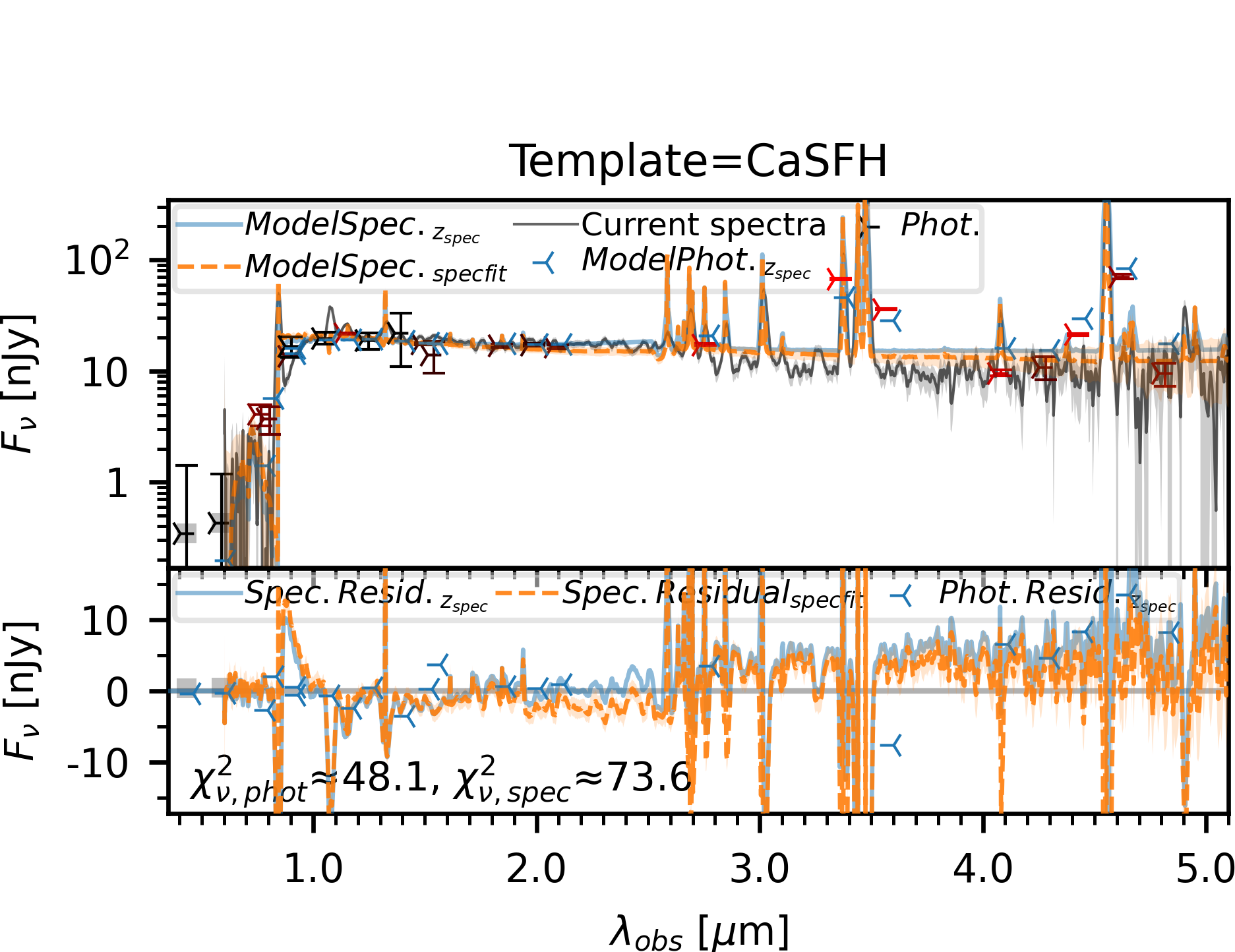}
    }
    \subfigure{
        \includegraphics[width=0.48\textwidth]{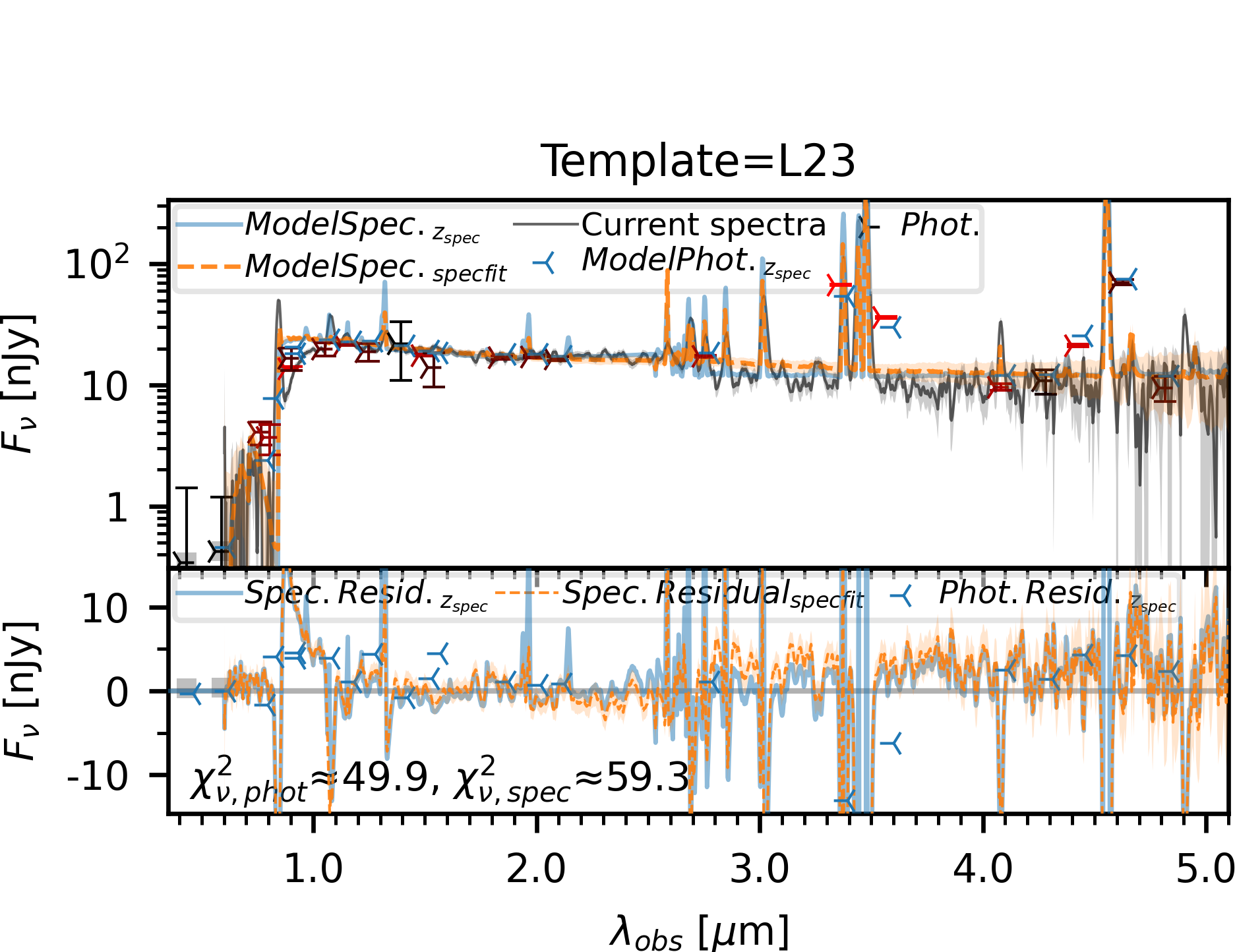}
    }
    \subfigure{
        \includegraphics[width=0.48\textwidth]{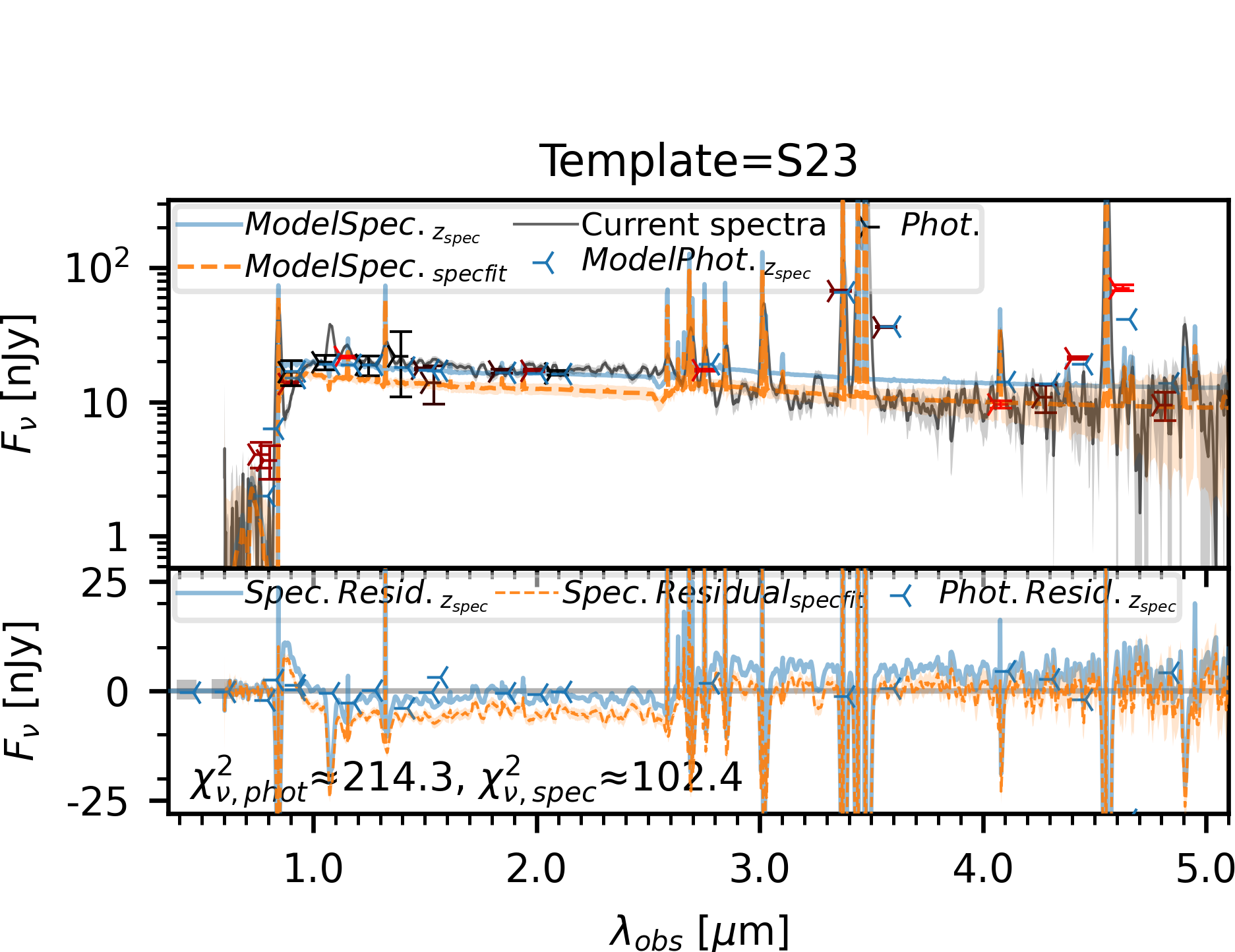}
    }
    \subfigure{
        \includegraphics[width=0.48\textwidth]{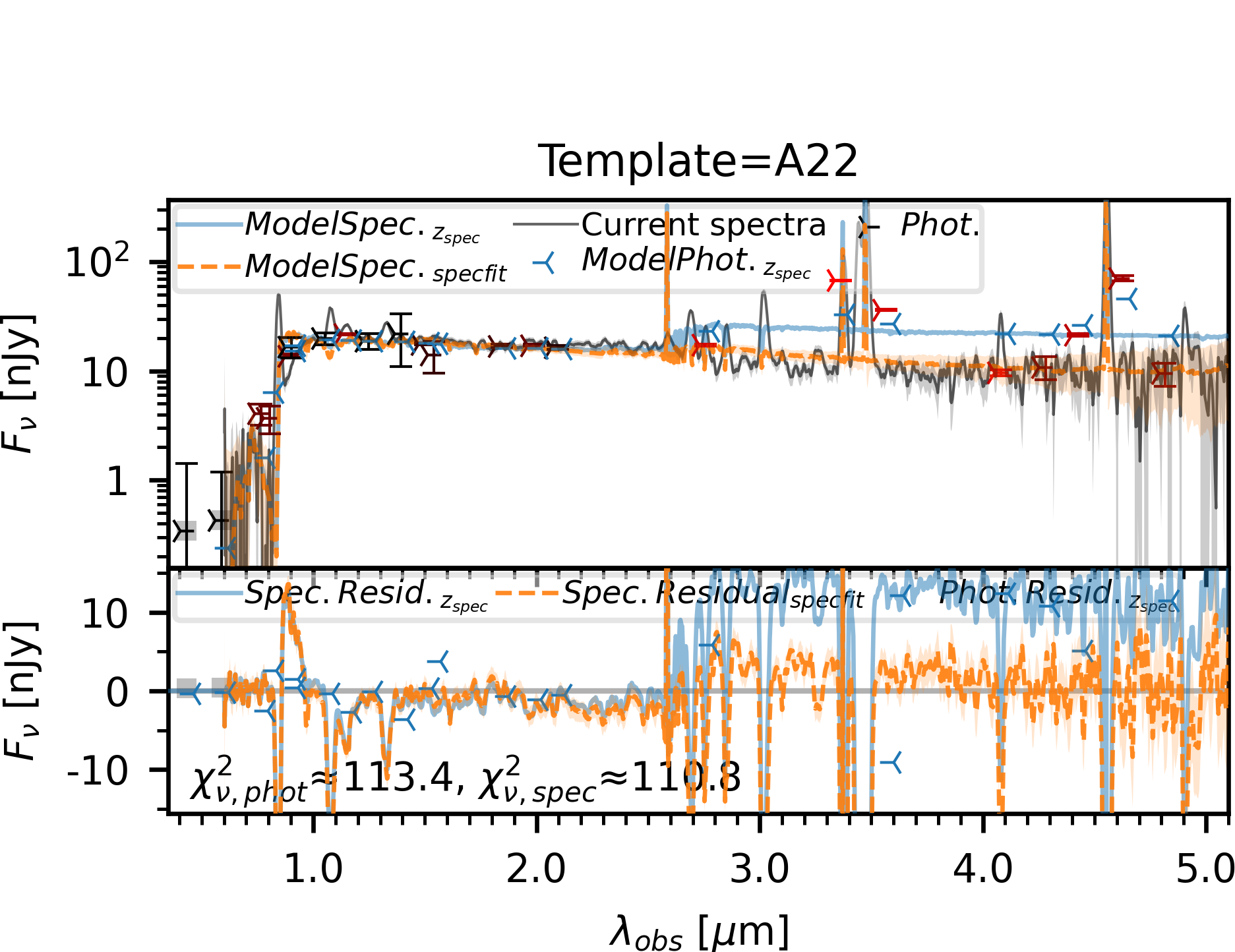}
    }
    \caption{\centering Best fits for object 113585 with various templates.}
\end{figure*}

\label{lastpage}
\end{appendix}
\end{document}